\documentstyle[11pt,aaspp4,flushrt]{article}

\newcommand\strain{\mbox{$\boldmath \Lambda$}}
\newcommand\vsh{v_{\rm sh}}

\def\be{\begin{equation}}
\def\ee{\end{equation}}
\def\k0{{{\pi e^2\over m c}f\chi\eta n}}

\newcommand{\phiorb}{{\phi_{\rm orb}}}
\newcommand{\phiprec}{{\phi_{\rm prec}}}

\newcommand{\partials}[2]{\frac{\partial #1}{\partial #2}}
\newcommand{\pfrac}[2]{\left(\frac{#1}{#2}\right)}

\newcommand{\Civ}{C{\sc iv}}
\newcommand{\Nv}{N{\sc v}}
\newcommand{\Siiv}{Si{\sc iv}}
\newcommand{\Heii}{He{\sc ii}}
\newcommand{\Ov}{O{\sc v}}

\newcommand{\Ostars}{\mbox{O\,stars}}

\newcommand{\vth}{v_{\rm th}}
\newcommand{\nhat}{\hat{n}}

\newcommand{\vvec}{\mbox{\bf v}}

\newcommand{\cmtwo}{\mbox{\,cm$^{-2}$}}

\newcommand{\kms}{\mbox{\,km~s$^{-1}$}}

\begin{document}

\title{Eclipsing Broad Emission Lines in Hercules X-1:\\
       Evidence for a Disk Wind?}

\author{James Chiang\altaffilmark{1,2}}

\affil{JILA, University of Colorado, Campus Box 440, Boulder CO 
       80309-0440}

\altaffiltext{1}{Present Address: NASA/Goddard Space Flight Center, 
                 Code 661, Greenbelt MD 20771}
\altaffiltext{2}{Also at Physics Department, University of Maryland, 
                 Baltimore County, Baltimore MD 21250}

\begin{abstract}
We present disk wind model calculations for the broad emission lines
seen in the ultraviolet spectra of the X-ray binary Hercules X-1.
Recent HST/STIS observations of these lines suggest that they are
kinematically linked to the orbital motion of the neutron star and
exhibit a red-shifted to blue-shifted evolution of the line shape
during the progression of the eclipse from ingress to egress which is
indicative of disk emission.  Furthermore, these lines are
single-peaked which implies that they may be formed in a disk wind
similar to those we have proposed as producing the broad emission
lines seen in the UV spectra of active galactic nuclei.  We compute
line profiles as a function of eclipse phase and compare them to the
observed line profiles.  Various effects may modify the appearance of
the lines including resonant scattering in the wind itself,
self-shadowing of the warped disk from the central continuum, and
self-obscuration of parts of the disk along the observer's
line-of-sight.  These latter two effects can cause orbital and
precessional phase dependent variations in the emission lines.  Hence,
examination of the line profiles as a function of these phases can, in
principle, provide additional information on the characteristics of
the disk warp.
\end{abstract}

\section{Introduction}

Hercules X-1 is one of the best-studied objects in the X-ray sky and
displays a wealth of behavior which make it a unique laboratory for
investigating the physics of accretion disks in binary systems.  It
emits X-ray pulsations with a 1.24\,s period indicating that the
compact object is a rapidly rotating neutron star whose magnetic field
is modulating the X-ray emission (Tananbaum et al.\ 1972).  In
addition, its hard X-ray spectrum shows features consistent with
resonant cyclotron absorption in a strong magnetic field, $B \ga
10^{12}\,$G, as would be expected at the neutron star surface
(Tr\"umper et al.\ 1978; Soong et al.\ 1990).  The system has a
1.7\,day orbital period, and the associated X-ray eclipses imply a
highly inclined ($i \ga 80^\circ$) circular orbit and an
intermediate-sized companion, HZ Herculis, with mass $M \simeq
2.2\,M_\odot$ (Deeter et al.\ 1981).  The X-ray light curve of Her X-1
is also modulated on a 35\,d period (Giacconi et al.\ 1973) and
consists of a ``main-high'' state lasting 10\,d, followed by a 10\,d
quiescent period, a subsequent ``short-high'' state of 7--8\,d, and
finally another quiescent period of 7--8\,d.  In addition, the X-ray
light curve shows pre-eclipse dips which occur at the beat frequency of
the orbital and 35\,d periods. The 35\,d cycle of Her X-1 and the
behavior of the pre-eclipse dips have been studied in great detail by
Scott \& Leahy (1999) using RXTE/ASM data.

The 35\,d modulation of the X-ray flux has been attributed to a
precessing accretion disk which is tilted and/or warped so that the
disk itself periodically obscures the central X-ray source (Katz
1973).  The origin of such a warp is not clear, but relatively recent
theoretical work has shown that both the warp and the disk precession
may be due to a radiation-driven warping instability (Pringle 1996;
Maloney, Begelman, \& Pringle 1996).  In these models, radiation from
the central source illuminates the disk and is absorbed.  Since this
central illumination is radially directed, it contributes no net
torque to any given annulus of the disk.  However, if the disk annulus
is optically thick, then the net pressure for re-radiated emission
will be normal to the surface of the disk.  Therefore, if the disk
annulus is illuminated non-axisymmetrically, a net torque can force
the annulus out of the nominal disk plane.  Non-axisymmetric
illumination may arise if the local angle of tilt of the disk varies
with position, i.e., if the disk is warped.  This is the basis for the
radiation-driven instability.  Alternatively, such torques may be
produced in a similar fashion by non-axisymmetric winds driven from
the disk surface by the X-ray heating (e.g., Schandl \& Meyer 1994;
Schandl 1996).  However, the nature of the underlying instability is
essentially the same.

Radiation-driven warping is an appealing mechanism since it allows for
global precessional modes which appear to be required by the
regularity of the long term periodicity, and it also provides a
natural time scale for the precession (Maloney, Begelman, \& Pringle
1996).  For Her X-1, a plausible estimate for this time scale can
easily account for the observed period: $\tau_{\rm prec} \sim
30\,(\alpha/0.1)^{-1}(\epsilon/0.1)^{-3}$ days, where $\alpha$ is the
usual disk viscosity parameter and $\epsilon = L/\dot{M}c^2$ is the
efficiency of the accretion (Maloney \& Begelman 1997).

The shape of the X-ray light curve over the 35\,d period can provide
some constraints on the characteristics of the disk warp and hence on
the physical mechanisms which produce it.  Schandl \& Meyer (1994) and
Schandl (1996) performed dynamical calculations of a disk warp driven
by coronal winds from the disk surface which are powered by heating
due to the central X-rays.  This model accounts for the presence of
the two high states via a combination of obscuration by the disk
itself and by an optically thick corona overlying it.  These authors
did not consider the effect of re-radiation torques in their
calculations, which may be at least as important as the torques due to
a wind (see discussion in Maloney, Begelman, \& Pringle 1996).  In
contrast, Scott, Leahy, \& Wilson (2000) have recently described a
completely phenomenological model for the tilt and phase of the inner
and outer disk annuli which are adjusted to fit the main-high and
short-high intervals.  In addition to differing from the Schandl-Meyer
geometry, their model accounts for the X-ray spectral softening seen
during both high state turn-ons as being due to absorption by cold
material in the outer disk.  Since the Schandl-Meyer model attributes
the short-high state to emergence of the X-ray source from behind the
hot, optically thick disk corona, it cannot account for this spectral
variation.  Although the X-ray observations can clearly place some
constraints on the nature of the putative disk warp, additional
observational diagnostics would be helpful.

In addition to driving warps, X-ray illumination of the disk can also
produce emission line radiation.  Several authors have performed
calculations of the disk structure and radiative transfer for an X-ray
illuminated disk (Raymond 1993; Ko \& Kallman 1994).  These models
predict the existence of an accretion disk corona as well as a
chromospheric layer which produces emission from a variety of lines
ranging from the Fe K lines in the X-rays to UV lines such as \Civ,
\Nv, \Siiv, \Heii, and \Ov.  Although the predicted line ratios appear 
to agree substantially with those observed in well-studied objects 
such as Her X-1 and Sco X-1, since they are produced in a disk, the 
lines are expected to be double-peaked, contrary to observations 
(Vrtilek et al.\ 1991; Boroson et al.\ 1996, 2000a).

We present a model for the broad single-peaked UV emission lines in
which they are formed in an accretion disk wind.  The underlying
mechanism is the same as the one we have proposed for the broad lines
in active galactic nuclei (Chiang \& Murray 1996; Murray \& Chiang
1997) and the single-peaked emission lines in cataclysmic variables
(Murray \& Chiang 1996): Large radial shears, due to the wind, in line
emitting material at the near and far sides of the accretion disk
allow line photons with small Doppler shifts to escape and produce
single-peaked emission lines.  The remainder of this paper is
organized as follows: In \S~2, we review the recent observations of
the UV emission lines in Her X-1, and in particular, discuss the
evidence for a disk origin for the broad UV lines.  In \S~3, we review
the mechanism for producing broad single-peaked emission lines in disk
winds.  In \S~4 we apply these calculations to fit the July 1998
HST/STIS data obtained during the eclipse phases of Her X-1.  We also
show how disk warping can affect the appearance of the lines as a
function of orbital and precessional phase.  Finally, in \S~5, we
present our conclusions.

\section{UV Observations}

Her X-1 has been observed several times in UV wave bands.  Spectra
were taken with the IUE satellite (Dupree et al.\ 1978; Gursky et al.\
1980; Howarth \& Wilson 1983), with the EUVE satellite (Leahy \&
Marshall 1999) and with IUE and EUVE during a multiwavelength campaign
involving the satellites ASCA, ROSAT, and CGRO (Vrtilek et al.\ 1994).
In 1994, high resolution UV spectra were obtained over nearly half of
the 1.7\,d orbital phase using HST/GHRS (Boroson et al.\ 1996) during
the main-high state.  Later, spectra were taken in July 1998 with
HST/STIS over almost a full orbital period, including eclipse ingress
and egress phases, while Her X-1 was in the short-high state (Boroson
et al.\ 2000a).  Most recently it was observed again by STIS in July
1999, but this time the source was in the ``anomalous low'' state in
which it was X-ray quiet even during nominal main-high and short-high
periods (Vrtilek et al.\ 2000).

From the 1994 GHRS observations, Boroson et al.\ (1996) reported
detecting broad high ionization lines \Heii, \Civ, \Siiv, and \Ov\ and
were able to resolve both narrow ($\delta v \sim 150$\kms) and broad
($\delta v \ga 700$\kms) components in \Nv.  They measured the Doppler
shifts of the broad and narrow components at the various orbital
phases by averaging the velocities at half-maximum of the line fluxes
and found that the broad and narrow components were correlated with
the expected Doppler shifts of the neutron star and companion
respectively.  In addition, they found that the narrow components
decreased in strength relative to the broad components as the system
approached eclipse of the primary.  They concluded that the narrow
line emission was produced on the surface of the X-ray irradiated
companion and that the broad emission was produced in the disk
surrounding the neutron star primary.

The July 1998 observations by Boroson et al.\ (2000a; hereafter B00)
using HST/STIS consist of 29 low resolution spectra obtained with the
G140L grating having a resolution of 200--300\kms\ and covering
wavelengths 1150--1720\AA, and 10 high resolution echelle spectra
obtained with the E140M grating with resolution 6\kms\ and covering
wavelengths 1150--1710\AA.  B00 confirmed the presence of both broad
and narrow components in \Nv\ and revealed similar features in the
\Ov, \Civ, \Siiv, and \Heii\ lines as well.  Just as with \Nv, the
narrow components of these other lines were not apparent during the
orbital phases when the X-ray illuminated face of HZ Her was not
visible, i.e., during the eclipse ingress and egress phases.  In the
absence of the narrow components, examination of the evolution of the
broad emission lines during the eclipse is particularly instructive.
Using a simple model for the obscuration of the disk by the companion,
B00 found that the fluxes in the broad \Nv\ and \Civ\ lines during the
ingress phases were consistent with being from a disk which produced
this emission symmetrically with respect to the line-of-centers
joining the primary and companion.  Furthermore, the high spectral
resolution data at the egress phase $\phiorb = 0.057$ showed that the
lines were blue-shifted as would be expected for a partially obscured
disk which rotates prograde with the binary orbit.

B00 did find some apparent difficulties with the disk interpretation
for the origin of the broad emission lines.  Although the Doppler
shifts of the fitted broad components of the \Civ, \Heii, and \Siiv\
lines were found to be largely consistent with the expected projected
velocities of the neutron star, the Doppler shifts of the fitted broad
\Nv\ components were red-shifted with respect to the neutron star.
This behavior is also seen in the measured velocities of the \Nv\ line
in the earlier 1994 GHRS spectra which were taken during the main-high
state.  In order to characterize this emission empirically, B00
modeled the disk emission from \Nv\ by dividing the disk into four
sectors corresponding to the left, right, front and back portions of
the disk.  They fit the high resolution spectra obtained during
eclipse egress of the July 1998 data and found that the red side of
the disk was brighter than the blue side by a ratio of at least 3:1.
Fits to the \Siiv, \Heii, and \Ov\ data have different angular
brightness distributions, but nonetheless do show evidence for
somewhat brighter red-shifted sides of the disk.  Notably, the red-blue
flux ratios found for the July 1998 short-high data also provided
adequate fits to the \Nv\ lines in the 1994 main-high GHRS data when
the disk was at a different precessional phase and was therefore
presumably at a different orientation.

\section{Broad Emission Lines from Accretion Disks}

The mechanism for producing single-peaked broad emission lines in a
disk-wind has been covered in some detail by Murray \& Chiang (1997,
hereafter MC97) where it was applied to the broad lines of AGNs and by
Murray \& Chiang (1996; see also Chiang \& Murray 1997) where it was
discussed in the context of broad single-peaked Balmer lines in
cataclysmic variable systems.  In this section, we present the
relevant physics and refer the interested reader to the above
references, particularly MC97, for more details.

From the observations of broad absorption line (BAL) QSOs and in the
context of the disk wind model, it is clear that the broad emission
line photons are produced via collisional excitation and/or
recombination.  Accordingly, the bulk of the line emission will
originate in the highest density regions of the wind near the base of
the streamlines where the wind proper interfaces with the disk (Murray
et al.\ 1995).  In the absence of a wind, the emission lines from a
Keplerian disk viewed at sufficiently large inclination will be
double-peaked.  If the emission is optically thin, this double-peaked
appearance is due to the zeroes of the azimuthal derivative of the
projected disk velocity occurring at the projected velocity maxima
(cf.\ Rybicki \& Hummer 1983; MC97).  Here, the projected disk
velocity is $v_{\rm proj} \equiv \vvec\cdot\hat{n}$ where $\hat{n}$ is
the unit vector in the direction of the observer and $\vvec$ is the
velocity of the disk material as a function of position.

For an optically thick line, the double-peaked appearance is somewhat
different.  In this case, the velocity shears in the disk
significantly alter the optical depth which a line photon sees for a
given photon direction.  In directions along which the velocity shears
are large, line photons escape more readily as atoms in material along
the photon path will be Doppler shifted out of resonance.  Conversely,
for paths along which the velocity shears are small, the atoms in the
material remain in resonance, the optical depths are large, and hence
line emission in these directions is suppressed.  In the near and far
sides of a Keplerian disk, relative to the inclined observer, where
the projected velocities of the disk material are near zero, the
velocity shears are also small.  This leads to suppression of emission
at wavelengths near line center.  Rybicki \& Hummer (1983) noted this
effect, and Horne \& Marsh (1986) showed that it provides a good
description of the Balmer line shapes seen in the optical spectra of
certain cataclysmic variables.  The presence of a disk wind with a
significant velocity shear in the poloidal direction will allow the
photons with small projected velocities from the front and back sides
of the disk to escape more easily, producing a single-peaked emission
line.

More formally, the local anisotropy of the line emission occurs
because the line emissivity is roughly proportional to the local line
width which in turn depends upon the anisotropic shear broadening.
The local apparent line surface brightness is given by
\begin{equation}
J = \cos i\, S \int d\nu\, \left[1 - \exp(-\tau_\nu)\right], 
    \label{local_emissivity}
\end{equation}
where $i$ is the inclination and the factor $\cos i$ accounts for
foreshortening assuming a flat disk, and $S$ is the line source
function.  The optical depth is
\begin{equation}
\tau_\nu = \frac{1}{\cos i}\int_{z_e}^\infty dz\, k_0 \phi_\nu(z),
\end{equation}
where $k_0$ is the line opacity which depends on the strength of the
transition and on the abundance of the relevant ion in the appropriate
state; in general, $k_0$ will vary with both $r$ and $z$.  The
function $\phi_\nu$ is the thermally broadened line profile shape
which has been Doppler shifted to the line-of-sight velocity of the
emitting material.  We assume a Gaussian shape for the profile:
\begin{equation}
\phi_\nu = \frac{1}{\sqrt{2\pi}\Delta \nu_{\rm th}}
           \exp\left(-\frac{1}{2}\left[\frac{\nu - \nu_0(1 + 
           \vvec \cdot \nhat/c)}
           {\Delta \nu_{\rm th}}\right]^2\right).
\end{equation}
The thermal line width is $\Delta\nu_{\rm th} = (\vth/c)\nu_0$ where
$\vth$ is the thermal velocity of the emitting gas.  We approximate
the line-of-sight velocity by expanding linearly from the base of the
emission layer at vertical height, $z_e$:
\begin{eqnarray}
\vvec \cdot \nhat 
   &\simeq& v_{\rm disk} + \frac{z - z_e}{\cos i} 
            \nhat\cdot\strain\cdot\nhat \nonumber\\
   &\simeq& v_{\rm disk} + \frac{z - z_e}{l_{\rm em}}v_{\rm sh}.
\end{eqnarray}
Here $l_{\rm em}$ is the vertical thickness of the emitting layer,
$v_{\rm disk} \simeq - v_\phi\sin i\sin\phi$ is the line-of-sight
velocity at the base of the emission layer, $v_\phi = \sqrt{GM/r}$ is
the azimuthal velocity, $\phi$ is the azimuthal angle of the emission
point, $v_{\rm sh}$ is the shear velocity along the line-of-sight
$\nhat$ (i.e., the velocity difference between the top and bottom of
the emission layer along the line-of-sight), and $\strain$ is the
symmetric strain tensor describing the flow.  In Cartesian
coordinates, the strain tensor is given by
\begin{equation}
\Lambda_{ij} = \frac{1}{2} \left(\partials{v_i}{r_j} 
               + \partials{v_j}{r_i}\right)
\end{equation}
The coordinates are defined with the disk symmetry axis coincident
with z-axis and the observer line-of-sight to the neutron star lying
in the xz-plane. We define the quantity $Q =
\nhat\cdot\strain\cdot\nhat$ which is the velocity gradient, or shear,
along the line-of-sight.

From Eq.~\ref{local_emissivity}, we see that for an optically thick
line, the local emissivity is approximately equal to the source
function multiplied by the width of the frequency interval over which
the optical depth exceeds unity.  Given the above expressions, the
local emissivity is then
\begin{equation}
J \simeq \cos i\, S \,\Delta\nu \sqrt{8 \log\tau_0}.
\end{equation}
The quantity $\tau_0$ is the central optical depth of the line,
\begin{equation}\label{tauzero}
\tau_0 = \frac{k_0 c}{\sqrt{\pi}\nu_0}
         \left[|Q|\sqrt{1 + \left(\frac{\vth}{\vsh}\right)^2}\right]^{-1},
\label{tau0}
\end{equation}
and the effective local linewidth is 
\begin{equation}
\Delta\nu = \Delta\nu_{\rm th} \sqrt{1 + \left(\frac{\vsh}{\vth}\right)^2}.
\end{equation}
Thus, when the line-of-sight shears are large, the local emissivity is
enhanced.

Accordingly, a disk wind which has large velocity shears in the radial
direction will have enhanced emission from the low projected velocity
material from the near and far sides of the disk.  For AGN disk winds,
just as for O stars, we found that the radial component of the wind
velocity can be modeled approximately as
\begin{equation}
v_r(r) = v_\infty\left(1 - \frac{r_{\rm f}}{r}\right)^{\gamma},
\label{vr_law}
\end{equation}
where $v_\infty$ is the terminal velocity of the wind, and $r_{\rm f}$
is the ``footpoint'' radius of the wind streamline (Murray et al.\
1995).  For O stars, this radius is that of the stellar surface; while
for a disk wind, it is the disk radius from which a given streamline
originates.  The velocity law exponent $\gamma$ depends on the
effective gravity and on the parameters describing the radiative
acceleration.  In both O stars and AGN disk winds, the bulk of this
acceleration is due to line driving.  The effectiveness of the lines
in providing driving force for the wind will be determined by the
ionization state of the wind material which will be photoionized by
either radiation from the O star or the AGN central continuum source.
Since the effective temperature of the inner regions of an AGN disk is
higher than that of an O star (${\cal O}(10^5)$\,K vs 40,000 K), the
line driving will be different.  In addition, because material in a
disk is in Keplerian motion, the effective gravity near the disk
surface is very small.  As a consequence, the velocity grows more
slowly near the disk surface than in the O star case.  In their
pioneering work on the dynamics of line-driven stellar winds, Castor,
Abbott, \& Klein (1975) derived a value of $\gamma = 0.5$, while later
work using more complete calculations of the line force found
$\gamma\simeq 0.8$ (e.g., Pauldrach, Puls, \& Kudritzki 1986).  In
Murray et al.\ (1995), we showed that the combined effects of the
reduced effective gravity and line force appropriate for an AGN
ionizing continuum, the velocity law exponent is $\gamma \sim 1.1$,
consistent with a slower radial acceleration near the wind footpoint.
Although the importance of line driving in accelerating a disk wind is
uncertain in Her X-1, the formalism we describe for the emission line
formation will be relevant for any radially accelerating disk outflow.

In a disk wind, for photons escaping along sightlines to a distant
observer, the line-of-sight velocity gradient is given by:
\begin{eqnarray}
Q &=& \sin^2 i \left[\partials{v_r}{r}\cos^2\phi + \frac{v_r}{r}\sin^2\phi
      + (3/2)\frac{v_\phi}{r}\sin\phi\cos\phi\right]\nonumber \\
  & & - \cos i\left[\partials{v_r}{r}(\sin i\cos\phi\cot\lambda + \cos i)
      + \frac{v_\phi}{2r\sin\lambda}\sin i\sin\phi\right]
\label{Full_Q}
\end{eqnarray}
where $\lambda$ is the angle at which the line-driven wind emerges
from the disk.  The monochromatic specific luminosity is 
\begin{equation}
L_\nu = \sum_{j = 1,2}\int r\,dr S(r) 
       \frac{l_{\rm em}}{v_\phi}
       \frac{|Q_j|\sqrt{1 + (\vth/\vsh)^2}}{\sin i|\cos\phi_j|}
       [1 - e^{-\tau_j}].
\label{line_luminosity}
\end{equation}
Here the source function $S(r)$ is assumed to depend only on radius.
The length scale of the emitting region is $l_{\rm em} = (v_{\rm
em}/v_\infty) r \sin\lambda$ where $v_{\rm em}$ is the velocity of the
emitting gas.  The subscript $j$ refers to the two azimuthal angles,
$(\phi_1,\phi_2)$, which correspond to the locations of the disk
material with Doppler shift of $(\nu-\nu_0)/\nu_0$ for a given radius
$r$.

As we noted above, the line emission is strongest where the density is
largest, near the footpoints of the wind.  Here the radial velocities
are only of order the thermal speed, $v_{\rm em} = v_r \simeq v_{\rm
th} \ll v_\phi$, and we can generally neglect the term proportional to
$v_r/r$ in Eq.~\ref{Full_Q}.  Furthermore, the vertical length scale
$l_{\rm em}$ will be much smaller than the disk radius.  In practice,
this implies that we can assume the emitting layer to be geometrically
thin and we can thus neglect any $z$-dependence of $k_0$ and $\tau$.
However, even though the radial velocities are small compared with the
azimuthal component, from our expression for $v_r$ (Eq.~\ref{vr_law}),
we see that the radial shears are larger: $\partial v_r/\partial r \ga
\partial v_\phi/\partial r$.  Hence, for high inclination objects, the
line-of-sight shear will be dominated by two terms:
\begin{equation}
Q \simeq \sin^2 i \left(\partials{v_r}{r}\cos^2\phi +
     \frac{3}{2}\frac{v_\phi}{r}\sin\phi\cos\phi\right).
\label{disk_wind_shear}
\end{equation}
The first term is the radial shear due to the wind, while the second
term is simply the Keplerian disk shear.  The effect of the radial
wind shear can be seen in the angular dependence of the first
term. Large radial shears at $\phi \simeq 0,\,\pi$, which correspond
to the near and far sides of the disk where the projected velocities
are small, allow line photons with small Doppler shifts to escape more
easily to the observer.

\section{Model Calculations and Comparison with Data}

We initially examine the simplest case of a flat disk, and we will use
the various eclipse light curve shapes as the principal constraints.
We model the eclipse of the disk by identifying the obscuring limb of
the companion with its Roche surface.  Assuming a primary mass of
1.3\,$M_\odot$, a companion mass of 2.2\,$M_\odot$, and an orbital
inclination of $i = 81^\circ\!.5$ (see e.g., Schandl 1996), the
primary will be eclipsed by the companion over orbital phases $\phiorb
= 0.936$--1.064.  If the accretion disk is flat and extends out to the
Roche radius, which is $r_{\rm Roche} = 2.6\times10^{11}$\,cm for
these system parameters, it will be at least partially eclipsed over
phases $\phiorb = 0.84$--1.16.  The maximum outer radius for the disk
to become fully eclipsed at $\phiorb = 1$ is $r_{\rm outer} =
2.4\times 10^{11}$\,cm.

Both Raymond (1993) and Ko \& Kallman (1994) found from their
calculations of X-ray illuminated accretion disks that the UV line
intensity from the disk surface should fall much less steeply than
$r^{-2}$ so that emission from the outer parts of the disk will
provide the dominant contribution to the line flux.  For optically
thick emission lines, the radial dependence of the line intensity will
be given by (see Eq.~\ref{line_luminosity})
\begin{eqnarray}
\frac{1}{r}\frac{dL_\nu}{dr} &\sim & S(r) \frac{l_{\rm em}}{v_\phi} Q \\
                             &\sim & r^{1/2} S(r).
\end{eqnarray}
In order to match the flat radial dependence of the line intensities
found by Raymond (1993; see his Fig.~4) and Ko \& Kallman (1994; see
their Fig.~18), we set $S(r) \propto r^{-1/2}$.  However, so long as
the outer disk radii make the main contribution to the line emission,
the exact radial dependence of $S(r)$ is not very important.  Since
the emission line shapes will also be largely insensitive to the
choice of inner radius of the line emitting region, for definiteness,
we fix the inner radius to be $r_{\rm inner}=10^9$\,cm, a value which
is comparable to the inner radii found by the above authors for the
lines we will examine.  For our fits to the emission line light
curves, it is sufficient to consider outer radii of the line emitting
regions in the range $r_{\rm outer} = 0.7$--$2 \times 10^{11}$\,cm
(cf. Table~\ref{best_fit_table}).  The lower limit for the outer
radius of the disk is the so-called circularization radius at which
the specific angular momentum of the disk equals that of the material
injected at the L1 point from the Roche lobe-filling companion (see,
e.g., Frank, King, \& Raine 1992), and an absolute upper limit for the
outer radius is the Roche radius.  However, for a given line,
particularly a high ionization one, the relevant outer radius may be
smaller than the actual disk radius since the emission properties are
also determined by the ionization state of the material.  In
principle, the appropriate value of the outer radius will be affected
by a number of factors including the widths of the lines, the residual
line flux seen at phases near mid-eclipse, and the shape of the
emission line eclipse light curve.

The eclipse spectra of the July 1998 HST/STIS observations presented
and analyzed by B00 constrain our model calculations.  Accordingly, in
order to match the resolution of the observations, we convolve our
model calculations with Gaussian functions having widths of 250\kms\
and 10\kms\ for the ingress and egress profiles respectively.  In
addition, since the high resolution data are oversampled, we rebin the
data appropriately.  The low resolution spectra were taken during the
ingress phases $\phiorb = 0.902$--0.986, and the high resolution
spectra were taken over the egress phases $\phiorb = 0.057$, 0.092 and
the out-of-eclipse phases $\phiorb = 0.132$--0.804.  As we noted
above, the out-of-eclipse spectra show prominent narrow emission lines
which B00 argue originate from the X-ray irradiated surface of the
companion. Since we do not model the narrow lines, we restrict our
attention to the low resolution ingress spectra and the first three
phases ($\phiorb = 0.057, 0.092, 0.132$) of the high resolution
spectra during which the narrow lines do not appear.  At each phase,
we compute emission line profiles and disk surface brightness maps.
For our flat-disk calculations, the precessional phases are
irrelevant; however, when we compute the effects of the warped disk
models of Schandl (1996) and Scott et al.\ (2000), the precessional
phases will be important.

\subsection{Eclipse Light Curves}
In Fig.~\ref{eclipse_lcs}, we plot the fluxes for the five lines,
\Heii, \Civ, \Nv, \Siiv, and \Ov, and the underlying UV continua
associated with each line as functions of orbital phase.  We also plot
our best-fit flat disk models for the eclipse light curves of the
lines.  For each of these lines, there is a small amount of residual
flux at phases during which the entire disk should be eclipsed by the
secondary.  For example, for the \Heii\ line, at orbital phase
$\phiorb = 0.985$, the residual line flux is $1.3 \times
10^{-14}$\,erg\,cm$^{-2}$s$^{-1}$ and is at approximately same level
at $\phiorb = 0.960$.  Unless the disk is fully eclipsed at both of
these phases, the coverage of the disk will be quite different, and
these line fluxes should not be the same.  Therefore, we conclude that
the residual flux at $\phiorb = 0.985$ does not consist of line
emission which we are seeing {\em directly} from the disk surface.
Instead, it may be line emission from the disk which has been
scattered either by resonant or continuum processes into our
line-of-sight by material high above the disk, such as the wind
itself.\footnote{I thank Bram Boroson for suggesting this hypothesis.}
Although the companion will still obscure some fraction of the wind
during the eclipse and therefore block part of the scattered emission,
for the purposes of fitting the eclipse profiles, we have subtracted
the mean of the small residual fluxes measured at $\phiorb = 0.984$
and 0.986 from the fluxes at all phases for each of the lines.

In fitting the eclipse light curves, we have allowed the disk outer
radius and overall disk flux normalization to vary, and we have used
our disk wind model along with the aforementioned source function law,
$S(r) \propto r^{-1/2}$, and inner radius, $r_{\rm inner} = 10^9$\,cm.
The other important model parameters are the ratio of the wind
terminal velocity to Keplerian disk velocity which we take to be
$v_\infty/v_\phi = 3$ (see Eq.~\ref{vr_law}), the exponent of the wind
radial velocity law which we set to $\gamma = 1$, and the optical
depth of the emission line.  The first two parameters determine the
relative magnitude of the radial and azimuthal gradients at the disk
surface.  However, as long as the radial and azimuthal gradients are
comparable, variations in these parameters will only have moderate
effects on the shapes of the line profiles and eclipse light curves.
Furthermore, since detailed modeling of the dynamics of the wind is
beyond the scope of this paper, we fix these parameters to these
exemplary values although both values may in fact vary with radius.
In addition, we fix the angle at which the wind emerges from the disk
to $\lambda = 30^\circ$.  As we note above, for high inclination
objects such as Her X-1, the line-of-sight shear will be largely
insensitive to this parameter (cf. Eq.~\ref{disk_wind_shear}).  We set
the poloidal component of the velocity to be $v_{\rm em} = 10$\kms\
which is approximately the thermal velocity for gas at a temperature
of $10^4$\,K.  Temperatures of this magnitude have been found for the
chromospheric layers in the calculations of Raymond (1993) and Ko \&
Kallman (1994).  The density of the emitting gas will depend on the
wind mass-loss rate.  In the context of this model, this density could
be inferred from the observations if the ionization state of the gas
were also known.  However, this would require a calculation similar to
that of Raymond and Ko \& Kallman which accounts for the effect of the
wind dynamics which could in turn depend on the gas ionization state
if radiative acceleration is important.

The line optical depth is more problematic.  As long as it remains
optically thick throughout the emission region, the line shape is
unaffected by the exact value since the $1 - e^{-\tau}$ factor in the
expression for the specific luminosity (Eq.~\ref{line_luminosity}) is
essentially unity.  However, if the line is optically thin, then $1 -
e^{-\tau} \sim \tau$, and any density dependence of the line opacity
$k_0$ will have an effect on the line shape (cf.\ MC97).  As we will
discuss below, marginally optically thick emission may be relevant for
the \Nv\ doublet which shows line ratios significantly different from
unity (B00) and which have a wider velocity profile than the other UV
lines we consider.

Since the \Heii\,1640 line is a singlet transition and does not show
evidence for interstellar absorption in any of the high resolution
spectra, it should be the easiest line to compare with the model
calculations.  We will therefore make it the primary focus of our
investigations.  The \Ov\,$\lambda 1371.292$ line is also a singlet
transition; but as B00 found, it appears to have either an absorption
component or a separate narrow emission component which will make its
spectra more difficult to model.  The \Siiv\ line is a doublet, but we
only measure the fluxes for the $\lambda 1393.755$ component since the
$\lambda 1402.770$ component is blended with a complex of O{\sc iv}
lines (B00).  Furthermore, the $\lambda 1393.755$ component is
affected by interstellar absorption.  However, we do consider the
\Civ\ and \Nv\ lines in some detail for our flat disk models, even
though they are doublets and are affected by interstellar absorption,
since their spectra allow us to examine some other effects which are
relevant for the properties of the wind and the chromospheric layer in
the disk, such as the implied wind column density in the case of the
\Civ\ lines and the effects of optical depth in the case of the \Nv\
lines.

In Fig.~\ref{eclipse_lcs}a, we plot the \Heii\ line fluxes as a
function of orbital phase.  These fluxes were obtained by fitting a
constant level to the UV continuum over wavelengths 1649--1663\AA\ and
integrating the residual line flux in the range 1631--1649\AA.  At
phases $\phiorb = 0.904$--0.913, there is a ``dip'' in the \Heii\
emission line light curve.  A similar dip is seen in the light curves
of \Siiv\ and \Ov.  These dips cannot be explained in the context of a
flat disk model.  One possibility, which we will explore below, is
that the appearance of the emission lines is modified due to
self-obscuration by a warped disk.  This will cause line shape and
flux variations which will be a function of precessional phase.  This
explanation also faces difficulties, however, since the eclipse light
curves of \Civ\ and \Nv\ do not show similar features at these phases.
In addition, the fits performed by B00 to the broad components of
these lines show that the fluxes vary differently for different lines
over the course of the full orbital period.  While some of these
variations may be due to uncertainties in fitting the narrow line
components simultaneously, they could reflect real differences in the
variations of the emissivity for the various lines.  It is unlikely
that this dip is related to the pre-eclipse dips which are seen in the
X-ray light curve.  Although, both phenomena occur at phases prior to
the main eclipse of the central X-ray source, the duration of the
X-ray dips is much longer and is comparable to that of the main
eclipse itself, $\Delta\phiorb = 0.13$, and it occurs at an earlier
orbital phase ($\phiorb \simeq 0.6$) than the dip which we see in the
emission line fluxes (Scott \& Leahy 1999).

In order account for the effect of these flux dips, we have fit the
eclipse light curves for two subsets of the data: a restricted
dataset, excluding the fluxes from phases corresponding to the dip
$\phiorb < 0.92$, and an unrestricted one, which includes all of the
data.  Fits to these two data subsets are shown as the solid and
dotted curves in Fig.~\ref{eclipse_lcs}a respectively.  For the
restricted dataset, we find a best-fit outer radius of $r_{\rm outer}
= 1.6 \pm 0.3\times 10^{11}$\,cm and an out-of-eclipse flux $F_{\rm
line} = 8.1 \pm 0.4 \times 10^{-14}$\,erg\,cm$^{-2}$s$^{-1}$.  For the
unrestricted dataset, the values of the outer radius and line flux we
obtain are similar: $r_{\rm outer} = 1.8 \pm 0.2\times 10^{11}$\,cm and
$F_{\rm line} = 7.1 \pm 0.2 \times 10^{-14}$\,erg\,cm$^{-2}$s$^{-1}$.
We will adopt the former values for our subsequent comparisons to the
individual \Heii\ line profiles.

In Fig.~\ref{eclipse_lcs}b, we show our fit to the
\Civ\,$\lambda\lambda$\,1548.195, 1550.770 light curve.  We find a
similar value of the outer disk radius for these data as our \Heii\
fit, and an out-of-eclipse flux of $2.2 \pm 0.1 \times
10^{-13}$\,erg\,cm$^{-2}$s$^{-1}$.  In contrast to the \Heii\ line,
the \Civ\ line flux at $\phiorb = 0.132$ is substantially larger than
the model value.  This cannot be accounted for by our flat disk
model. However, the best fit parameters are not significantly affected
by inclusion of this data point.  In addition, we have accounted for
the \Civ\ interstellar absorption mentioned by B00, which we
determined by fitting the post-eclipse $\phiorb = 0.132$ line profile.
These features can be clearly seen in the $\phiorb = 0.092$ profile in
Fig~\ref{Civ_profiles}.  The inclusion of these absorption components
in the model does not significantly change the light curve shape.

For comparison, Table~\ref{best_fit_table} shows the values of $r_{\rm
outer}$ and $F_{\rm line}$ which we have obtained for the various
emission lines.  It is worth noting that the higher ionization lines,
\Nv\ and \Ov\, have systematically smaller outer radii.  This is
similar to the so-called ``ionization stratification'' which is
inferred from reverberation mapping of the broad line regions of
Seyfert Galaxies.  For these objects, the lower ionization emission
lines are found to be produced at larger radii than the higher
ionization lines (e.g., Krolik et al.\ 1991; Korista et al.\ 1995).

\subsection{Line Profiles}
In Fig.~\ref{HeII_profiles}, we show, at orbital phases $\phiorb =
0.909$, 0.939, 0.057, 0.092, the apparent disk surface brightness,
including partial obscuration by the companion (left panels), the
observed and model spectra for the \Heii\ line (middle), and the
residual spectra after subtraction of the model line profiles (right).
Although we could, in principle, show comparisons of the model with
the data for all 29 spectra during the ingress phases, the spectra at
$\phiorb = 0.909$ and 0.939 are representative of the quality of the
model fits for the spectra during the phase intervals $\phiorb =
0.902$--0.916 and $\phiorb = 0.938$--0.960, respectively.  In order to
give the reader some idea of how well the models describe the
individual line profiles, we present in Table~\ref{chi2_values},
values of $\chi^2$ and numbers of degrees-of-freedom (dof) for the
various models and line profiles shown in Figures~\ref{HeII_profiles},
\ref{Civ_profiles}, and \ref{Nv_profiles}, as well as for the \Siiv\
and \Ov\ lines.  It is clear from Tables~\ref{best_fit_table} and
\ref{chi2_values} that the emission lines are not well described by
these models in any {\em formal} statistical sense.  That is beyond
the scope of this paper.  Instead, this work is meant to provide a
plausibility argument for the origin of these lines in a disk wind,
leaving to future work the various details of the model which are
required to produce statistically satisfactory fits to the data.

The surface brightness pattern shown in Fig.~\ref{HeII_profiles} will
be the same for all the lines since we assume the same velocity for
the emitting gas and the same radial dependence for the source
function $S(r)$ with only the outer radius differing from line to
line.  The ratio of the emitted fluxes for the brighter to the darker
regions at the outer radius is approximately a factor of 20.  At
$\phiorb = 0.909$, the observed line profile is red-shifted relative
to the model profile.  This appears as blue-shifted absorption in the
residual spectrum and also holds for each of the profiles for phases
$\phiorb = 0.902$--0.916.  One interpretation of this absorption is
that it is due to an out-flowing wind, such as the disk wind itself,
which is scattering the underlying continuum and line emission.  In
Fig.~\ref{early_phase_residuals}a, we plot the mean residual spectrum
covering the ingress phases 0.902--0.916 for the
\Heii\ line, except that instead of subtracting the mean residual line
fluxes at $\phiorb = 0.984,\,0.986$, we have instead added this
residual flux to our model.  We find a similar result for the \Civ\
lines, and we plot these residuals in
Fig.~\ref{early_phase_residuals}b.  For the fits to the \Civ\ lines,
we have again used the eclipse light curve to constrain the outer
radius of the \Civ\ emitting region and to find the normalization of
the out-of-eclipse line flux (Fig.~\ref{eclipse_lcs}b).  We also
assume that the lines are optically thick so that the doublet ratio is
unity.

In the context of the wind interpretation for these residual
P\,Cygni-like line profiles, we can place a rough lower limit on the
wind column density based on the amount of blue-shifted absorption
seen.  For velocities less than zero, the equivalent width of the
absorption in \Civ\ is $\sim 300$\kms.  In an expanding flow, at a
given frequency, the optical depth will be $\tau \simeq n \sigma l_s$,
where $n = n_H \chi_{\sc c} \eta_{{\sc c}^{+3}}$ is the number density
of absorbers, $\chi_{\sc c}$ is the abundance of carbon, $\eta_{{\sc
c}^{+3}}$ is the fractional abundance of the C$^{+3}$ ion, $\sigma$ is
the resonant-scattering cross-section, and $l_s = \vth/(dv_r/dr)$ is
the Sobolev length scale.  The width at zero intensity of the
absorption is $\sim 800$\kms, so taking a mean optical depth of $\tau
\sim 3/8$, we find a column density of $N_H \ga 10^{18} \eta_{{\sc
c}^{+3}}^{-1}$\cmtwo.  In making this estimate, we have also assumed
the terminal velocity for the absorbing part of the wind to be 800\kms
and that all of the carbon atoms are in the C$^{+3}$ ground state.
From this column density, we also estimate a wind mass loss rate of
$\dot{M}_{\rm w} \sim 10^{13} \eta_{{\sc c}^{+3}}^{-1}$\,g\,s$^{-1}$
using the radial velocity law, Eq.~\ref{vr_law}, assuming solar
abundance of carbon and a characteristic wind footpoint radius of
$r_{\rm f} \sim 10^{10}$\,cm.  These estimates of the wind column
density and mass loss rate are very crude, but the implied value of
$\dot{M}_{\rm w}$ is comfortably below the Eddington accretion rate
$\dot{M}_{\rm Edd} \sim 10^{17}$\,g\,s$^{-1}$.

\subsection{Marginally Optically Thick \Nv\ Emission}
Before turning to the possible effects of a warped disk on the shape
of the disk emission lines, we consider the \Nv\ emission.  As noted
by Boroson et al.\ 1996 and B00, the doublet ratios for the
\Nv$\lambda\lambda1238.821,\,1242.804$ lines are significantly
different from unity.  For the July 1998 data, the values range from
unity to 1.6.  In their models of X-ray illuminated accretion disks,
Raymond (1993) and Ko \& Kallman (1994), found that the UV emission
lines, including \Nv, should be optically thick and therefore that the
doublet ratio for these lines should be unity.  By contrast, in the
optically thin limit, the doublet ratio should be proportional to the
ratio of oscillator strengths.

As we noted in our discussion of \S~3, another indication of an
optically thin line is a double-peaked profile.  Optically thin lines
will also be broader at half maximum than a single-peaked optically
thick line.  To illustrate this, we show in Fig.~\ref{tau_profiles}
disk wind emission lines for which we have varied the absolute scaling
of the optical depth.  From Eq.~\ref{tau0}, the radial dependence of
the line optical depth to lowest order is contained in two factors,
the shear along the line-of-sight $Q \sim v_\phi/r \sim r^{-3/2}$
(cf. Eq.~\ref{disk_wind_shear}) and the line opacity $k_0 \sim \kappa
\rho$.  Here $\kappa$ is the absorption coefficient which depends on
the atomic properties and on the abundance of the relevant ion, and
$\rho$ is the mass density.  For simplicity, we will assume that the
bulk of the emission for a given ion coincides with the locations in
the disk wind for which the fractional abundance of that ion peaks.
In ionization equilibrium, the relative fraction of ion $i+1$ to ion
$i$ is proportional to the ionization parameter:
\begin{equation}
\frac{n_{i+1}}{n_i} \propto U \equiv \frac{n_\gamma}{n_{\sc h}},
\end{equation}
where $n_\gamma$ is the number density of hydrogen ionizing photons,
and $n_{\sc h}$ is the number density of hydrogen atoms.  Under the
aforementioned assumption, the absorption coefficient $\kappa$ is also
proportional to $U$, and we have $k_0 \propto n_\gamma \propto F_{\rm
inc}$ where $F_{\rm inc}$ is the incident flux from the central X-ray
source.  For a flat disk and at large disk radii, the incident flux
from a central source goes as $F_{\rm inc} \sim r^{-3}$.  However,
several effects, such as the possible flaring of the disk and downward
scattering of the flux from the central source by material above the
disk, can make the radial dependence of the incident flux closer to
$r^{-2}$.  In fact, both Raymond (1993) and Ko \& Kallman (1994) model
the incident flux using this radial dependence.  Therefore, in order
to be consistent with these authors, we take $k_0 \propto r^{-2}$;
consequently, the optical depth will have a $r^{-1/2}$ scaling.  In
our calculations, we model the line optical depth as
\begin{equation}
\tau = \tau_c \frac{1}{Q_{11}}\pfrac{r}{10^{11}\,{\rm cm}}^{-1/2},
\label{tau_scaling}
\end{equation}
where $Q_{11} \equiv Q/(v_\phi/r)_{r=10^{11}{\rm cm}}$\null.  In
Fig.~\ref{tau_profiles}, we have set $\tau_c = 10^6$ (dashed curve),
$3$ (solid), and $10^{-6}$ (dotted).  The $\tau_c = 3$, marginally
optically thick line produces a single-peaked, flat-topped profile
with a larger width at half-maximum than the optically thick line.

A line shape similar to the marginally optically thick line shown in
Fig.~\ref{tau_profiles} provides a better fit to the \Nv\ profiles
than the optically thick profile.  In Fig.~\ref{Nv_profiles}, we show
fits to the \Nv\ profiles using both $\tau_c = 3$ emission lines
(solid curves) and an optically thick line (dashed).  In the
marginally optically thick fits, we have adjusted the doublet ratio to
be 1.22 since this is the expected doublet ratio if the stronger
member of the doublet, \Nv\,$\lambda$1238.821, has an optical depth of
3.  This doublet ratio also provides a reasonable estimate of the
relative strength of the two components in the out-of-eclipse profile
at $\phiorb = 0.132$ (cf.\ B00).  As for the \Heii\ and \Civ, the
out-of-eclipse line flux normalization and outer radius of the
\Nv-emitting region was found by fitting the eclipse light curve.  For
the $\tau_c = 3$ calculations, we find $r_{\rm outer} = 1.0 
\times 10^{11}$\,cm and an out-of-eclipse line flux of $F_{\rm line} =
3.8 \pm 0.1 \times 10^{-13}$\,erg\,cm$^{-2}$s$^{-1}$, while for the
optically, thick model, we obtain $r_{\rm outer} = 1.4 \times
10^{11}$\,cm and essentially the same out-of-eclipse line flux.  We
have also performed these calculations using $\tau_c = 1.5$ and find
only small differences in the light curve and line profile shapes.  In
order to account for both the eclipse light curve and emission line
shapes in this model, these results suggest that \Nv\ line needs to be
marginally optically thick with an optical depth of $\tau \sim 1$--3
at $r = 10^{11}$\,cm.

\subsection{Emission Lines from a Warped Disk}
The P\,Cygni-like residual profile is notably absent for the ingress
phases $0.938 < \phiorb < 0.960$, during which the neutron star is
obscured (Fig.~\ref{HeII_profiles}).  This would be consistent with
absorption in a wind if the source of the UV continuum was centered on
the neutron star and much smaller than the disk.  However, the shape
of the eclipse of the UV continuum (Fig.~\ref{eclipse_lcs}f) is
similar to that of the lines implying that the UV emitting region is
also similarly extended.  Matters are complicated further by the fact
that the companion star may also make a significant contribution to
the UV continuum (Leahy \& Marshall 1999).  In any case, the absence
of P\,Cygni-like features during phases $\phiorb \ga 0.940$ is
problematic for the disk wind interpretation.  One explanation is that
a wind does not originate from every region of the disk, and hence the
portions which are unobscured during phases $\phiorb \ga 0.94$ do not
exhibit blue-shifted absorption since there is no wind along the
line-of-sight.  If this were the case, however, we would expect a line
profile shape more appropriate for a disk without a wind during these
phases, whereas the disk wind profiles fit the data reasonably well at
these phases.

An alternative explanation is that the apparent lack of blue-shifted
line emission during $\phiorb \la 0.916$ may be due to the warped
shape of the disk.  Just as obscuration of the neutron star by the
warp is invoked to explain the X-ray low states of the 35\,d cycle,
various parts of the line-emitting regions of the disk may also be
blocked from our line-of-sight.  In addition, self-shadowing of the
disk is expected to occur which will affect the dynamics of the
radiation-driven warping (Pringle 1997; Wijers \& Pringle 1999).  As
far as the line emission is concerned, regions which are not exposed
to the central X-ray source may not produce any line emission, or if
those regions are illuminated indirectly by X-rays scattered by the
wind (cf.\ Ko \& Kallman 1994; Shakura \& Sunyaev 1973), reduced
emission may occur.  In either case, the shadowed regions can in
principle have different emission characteristics than the
non-shadowed regions, and in both cases, line-of-sight obscuration by
the disk itself will play a role.

In order to examine the effects of a warped disk, we consider two
models for the shape of the warp in Her X-1.  Although neither of
these models is likely to be the correct one for describing the disk
warp, they will serve to illustrate the effect that a warp can have on
the line profile shapes.  The first model is that of Schandl \& Meyer
(1994) and Schandl (1996) in which the warp is generated by torques
from a Compton-heated wind.  Although, this model has been criticized
on both theoretical (Pringle 1996; Maloney, Begelman, \& Pringle 1996)
and observational (e.g., Scott et al.\ 2000) grounds, it provides a
useful illustration of the shadowing effects we wish to explore.  In
order to compute the disk shape for this model, we use the Eulerian
angles of the disk annuli shown in Fig.~1 of Schandl (1996).  The
tilts of the annuli range from 4$^\circ$ at the inner radius, passing
through a maximum of 8$^\circ$ at $r \simeq 10^{11}$\,cm, and having a
value of $\la 7^\circ$ at the outer radius of $1.7\times 10^{11}$\,cm.
The phases of the lines of nodes range from zero at the inner radius
to $\simeq -270^\circ$ at the outer radii.

The second model we consider is the phenomenological one proposed by
Scott et al.\ (2000; hereafter SLW), based upon the shape of the 35\,d
light curve.  Here, only the tilts and precessional phases of the
inner and outer edges of the disk are specified.  The tilts of the
inner and outer disk annuli are 11$^\circ$ and 20$^\circ$ respectively
with a relative phase shift of the lines of nodes of these two annuli
of $138^\circ\!.6$.  In the absence of further information, we simply
assume that the Euler angles describing the disk warp vary linearly
with radius between these two limits.  In this model, emergence of the
X-ray source from behind the distant outer disk edge is responsible
for the sharp turn-ons, while the gradual declines at the end of each
high state is due to progressive occultation by the inner disk edge
which is sufficiently close to the X-ray source so that it takes a
significant amount of time to cover it completely.  From examination
of the occultation of the 1.24\,s X-ray pulsations, SLW infer an inner
disk radius of $r_{\rm inner} = 20$--40 neutron star radii.  For
definiteness we set $r_{\rm inner} = 4\times 10^7$\,cm.  SLW also
assume an observer inclination of $i = 85^\circ$.  It is interesting
to note that their model requires that the X-ray emission from the
neutron star comes from different hemispheres depending on whether Her
X-1 is in the main-high or short-high states.  For the emission line
calculation, because of the large observer inclination and large disk
tilts in their model, this means that emission line radiation will be
visible from {\em both} sides of the accretion disk, assuming it is
geometrically thin.  This is in contrast with the Schandl-Meyer model
in which only one side of the disk and X-ray source are visible.

We show in Fig.~\ref{warp_light_curves} light curves for these two
models assuming as before a line source function which depends only on
radius $S(r) \propto r^{-1/2}$.  The major difference between these
calculations and the flat disk models is that we account for
obscuration along the line-of-sight due to the disk warp.  In
addition, because each disk annulus has a radius-dependent tilt, the
velocity field of the flow and the Doppler shifts of each
line-emitting disk wind element will be modified from the flat disk
case.  Since the disks are warped, the precessional phases are now
important.  For the 1998 July observations, B00 found $\phiprec =
0.64$--0.74 corresponding to the short-high state using the high-state
turn-on measurements of Scott \& Leahy (1999).  We note that the
ingress and egress data of B00 were separated by a full orbital
(1.7\,d) period.  This partially accounts for the large difference
between the model line fluxes before and after the eclipse.  As
expected, the magnitude of this difference depends on the specific
shape of the disk warp.

The Schandl-Meyer model (Fig.~\ref{warp_light_curves}a), specifies the
outer disk radius as $r_{\rm outer} = 1.7 \times 10^{11}$\,cm which
for simplicity we leave fixed at this value, although the size of the
emitting region can in principle vary depending on the line
considered.  As with the flat disk model, we fit the \Heii\ line
fluxes over two orbital phase ranges, one excluding the dip $\phiorb =
0.920$--1.132 (solid curve) and the other over the full range
0.902--1.132 (dashed).  For the SLW model
(Fig.~\ref{warp_light_curves}b), only the tilt of the outer radius is
specified and we allow both the line normalization and this outer
radius to vary, finding a best-fit value of $r_{\rm outer} = 1.8 \pm
0.2\times 10^{11}$\,cm.  Unlike the Schandl-Meyer model, fitting both
ranges of orbital phase for the \Heii\ data using the SLW model gives
nearly identical results.  This is due to the particular shape of the
light curve produced by the SLW model which has a less steep decline
in flux during the ingress phases than the light curves produced by
either the flat or Schandl-Meyer models.  On egress, the SLW model
light curve is similar to that of the flat disk model, whereas the
Schandl-Meyer model yields an eclipse light curve with steeper ingress
and egress phases than those of the other two models.

In Figs.~\ref{SM_warp_line_profiles} and~\ref{SLW_warp_line_profiles},
we show the apparent disk surface brightness for the Schandl-Meyer and
the SLW models, and the line profiles associated with each model
compared with the \Heii\ data.  Although, the Schandl-Meyer model
shows a great deal of obscuration by the disk, it is largely red-blue
symmetric so that the model profile is still noticeably blue-shifted
at $\phiorb = 0.909$.  By contrast, the SLW model at this phase has a
larger portion of its disk obscured on the approaching side of the
disk and therefore the overall Doppler shift of the line is in better
accord with the data.  At high spectral resolution ($\phiorb =
0.057,\,0.092$), the differences between the two models are much more
apparent.  The SLW profiles are narrower at zero intensity, owing
mostly to the larger outer disk radius, while the Schandl-Meyer
profiles exhibit more structure due to its more complex obscuration
pattern on the disk surface.

\section{Conclusions}

We have presented calculations of the broad UV emission line profiles
which are formed in an accretion disk wind for the X-ray binary Her
X-1.  Calculations performed assuming a flat accretion disk reproduce
the eclipse light curve shapes for the July 1998 HST/STIS data
reasonably well, although previous analysis of these data revealed
out-of-eclipse variability in the broad line fluxes which are probably
due to time-dependent variations in the emitting properties of the
disk material.  The flat disk line profiles also do a reasonable job
of reproducing the shapes of the emission lines during eclipse as they
evolve from red-shifted on ingress to blue-shifted on egress.

The flat disk model does not provide a perfect description of the line
shapes however.  At orbital phases $\phiorb = 0.902$--0.916, the
observed profiles are more red-shifted than the model calculations.
This may be due to scattering in an outflowing wind as evidenced by
the residual P\,Cygni profiles for these phases
(Fig.~\ref{early_phase_residuals}).  This explanation is attractive
since P\,Cygni line profiles appear in the 1999 July HST/STIS spectra
of Her X-1 at orbital phases $\phiorb \sim 0.5$ (Boroson, Kallman, \&
Vrtilek 2000b).  Alternatively, the relative red-shifts of the
observed line shapes may be due to self-obscuration of the
blue-shifted portions of a warped disk
(Figs.~\ref{SM_warp_line_profiles} and~\ref{SLW_warp_line_profiles}).
Such self-obscuration may also be able to account for the line shape
differences which are seen in the high resolution spectra at $\phiorb
= 0.057$ which the flat disk model consistently overpredicts
(Figs.~\ref{HeII_profiles}, \ref{Civ_profiles}, \ref{Nv_profiles}).
However, B00 note that there is evidence for narrow absorption in the
\Nv\ and \Civ\ profiles at a Doppler shift of $-500$\kms\ which may
significantly modify those line profiles at $\phiorb = 0.057$ since
the overall Doppler shifts of those lines are also $\sim -500$\kms\ at
this phase.  This absorption also appears at phases $\phiorb = 0.092$,
0.132, and 0.171 at the same velocity, so it is unlikely to be
associated with the disk either as an obscuring warp or outflow, as
the overall Doppler shift would then vary by $\sim 100$\kms\ over
these phases (B00).

The main difficulty which this model faces is how a wind with the
required characteristics can be launched from the surface of the disk.
In previous analytic and numerical calculations of disk winds in AGNs
and CVs, line driving provided the mechanism for accelerating the wind
in which most of the initial acceleration was due to radiation
produced in the disk itself (e.g., Murray et al.\ 1995; Proga, Stone,
\& Drew 1998).  Although our treatment does not specifically require
radiative acceleration to play a substantial role in producing the
wind, large poloidal velocity gradients in the emission line region
are a necessary ingredient.  In the case of X-ray binaries such as Her
X-1, it has been posited that any wind emerging from the disk would
likely be thermally driven and have its origins in a Compton-heated
corona which has been irradiated by the central X-ray source
(Begelman, McKee, \& Shields 1983).  The scale height for such a wind
should be comparable to the disk radius.  This violates the underlying
assumption in these calculations of a geometrically thin, albeit
warped, disk.  Furthermore, given these large vertical scale heights,
it is unlikely that there will be poloidal velocity gradients which
are comparable to the Keplerian velocity gradients.

An alternative mechanism for driving a wind from a disk is via
magneto-centrifugal acceleration (e.g., Blandford \& Payne 1982).  In
this model, the magnetic field lines thread the accretion disk and are
anchored to the disk by flux freezing.  If the field lines rotate
rigidly with the disk and make an angle greater than $60^\circ$ with
the disk normal, then material will be centrifugally accelerated along
the field lines.  If the magnetic field is largely poloidal then the
acceleration of the disk material can be estimated by the condition
that the angular velocity is constant along the field line.  This
leads to radial shears which are comparable to the Keplerian shears in
the disk and single-peaked emission lines can be produced (Chiang \&
Murray 1997).  However, in the self-similar solution of Blandford \&
Payne (1982; see also Emmering, Blandford, \& Shlosman 1992), the
magnetic fields near the disk are not mostly poloidal since the
inertia of the accelerated material actually distorts the field lines
and causes them to have a significant toroidal component in the
$-\hat{\phi}$-direction.  The azimuthal shears are actually enhanced;
the radial shears are much smaller than they would be if the field
lines were poloidal; and as a result, the emission line profiles are
strongly double-peaked (Chiang 2000).

It is far more likely in the case of Her X-1 that radiative
acceleration does play a role in the formation of the broad emission
lines.  The existence of a chromospheric layer in which atoms with
strong resonance lines are abundant implies that line driving pressure
may well be important in determining the vertical structure of a X-ray
irradiated disk atmosphere.  For \Ostars, Castor, Abbott,
\& Klein (1975; hereafter CAK) argued that even in a static atmosphere
the force due to lines will greatly exceed that due to gravity, making
a line-driven wind inevitable.  A cursory examination of the
properties of the X-ray irradiated disk atmospheres computed by
Raymond (1993) and Ko \& Kallman (1994) reveals that similar
conditions exist which may allow significant line-driving to occur.

Following CAK, we use a force multiplier formalism to describe the
effects of line driving. Here the enhancement factor relative to
electron scattering, the force multipler ${\cal M}$, is modeled as an
inverse power of the electron scattering optical depth: ${\cal M} = k
t^{-\alpha}$, where $k$ and $\alpha$ are parameters which depend on
the driving continuum and on the composition and ionization state of
the irradiated gas.  For a static atmosphere, the electron scattering
optical depth is $t \simeq \sigma_e\rho l$ where $\sigma_e$ is the
electron scattering opacity, $\rho$ is the mass density, and $l$ is
the appropriate length scale which, for a disk wind, is $l_{\rm em}$.
In the case of a X-ray irradiated disk atmosphere, we can estimate the
relevant value of $t$ from the line optical depths quoted by Raymond
and Ko \& Kallman.  The latter authors find line optical depths of
$\ga 10^4$.  For resonance lines such as \Civ\ this implies an
electron scattering depth of $t \sim 10^{-5}$ for the chromospheric
layer assuming a gas temperature of $10^4$\,K.  For \Ostars, such a
small value of $t$ corresponds to the regime where ${\cal M}$ has
saturated at ${\cal M}_{\rm max} \sim 10^3$ essentially because of the
oscillator strength sum rule (CAK; Abbott 1982).  At disk radii $r
\sim 10^{10}$\,cm, Ko \& Kallman (1994) find disk photosphere
temperatures in the range $T_{\rm disk} =10^4$--$10^5$\,K, which is
comparable to that of \Ostars\ making our use of this saturation value
reasonable.  Using this value of the force multiplier and $F = \sigma
T_{\rm disk}^4$ for the driving flux, the acceleration due to
radiation pressure from the disk is $f_{\rm rad} = \sigma_e F {\cal
M}/c \simeq 6\times 10^5 (T_{\rm disk}/3\times10^4{\rm
K})^4$\,cm\,s$^{-2}$.  This exceeds the vertical component of gravity
$f_g \simeq 10^5$\,cm\,s$^{-2}$ assuming a central mass of
1.3\,$M_\odot$, a disk radius of $10^{10}$\,cm and a vertical height
for the chromosphere of $10^9$\,cm.

These arguments suggest that it would be worthwhile to perform
calculations of X-ray irradiated disk atmospheres which include the
effects of radiative acceleration.  Such calculations would likely
result in substantial changes to the disk vertical structure, line
optical depths, and local energy balance, and may be able to account
for the appearance of marginally optically thick emission lines such
as those we have inferred for \Nv.  Furthermore, if radiatively driven
winds do result, they may be important for producing a disk warp.  For
a given incident flux of radiation, winds can be more efficient at
driving warps than radiation pressure (Pringle 1996).  However, given
the large scale heights of thermally driven winds such as those
examined by Schandl \& Meyer, it is difficult to understand the manner
in which the wind back pressure is communicated down to the disk plane
in order to produce non-axisymmetric torques on the underlying disk
annuli.  By contrast, the vertical length scale of a radiatively
driven wind will be much smaller than that of a thermally driven wind
since the gas does not have to reach the escape temperature in order
to become unbound.

So it seems plausible that radiative acceleration plays a role in the
dynamics of the line emitting regions of the disks in X-ray binaries.
The appearance of broad single-peaked lines in the UV spectra of Her
X-1 and the abundant evidence from the eclipse light curves and line
profiles implying a kinematic connection with an accretion disk makes
this scenario even more attractive.

\acknowledgements 

I thank Bram Boroson for providing data from the 1998 July HST/STIS
observations and for several useful discussions.  I thank Norm Murray
for helpful comments and useful discussions.  I also acknowledge
helpful comments by the anonymous referee which have improved this
paper.  This work was support by NASA ATP grant NAG 5-7723.  In
conducting this research, the NASA Astrophysics Data System (ADS)
Abstract and Article services have been used.

\begin{table}[p]
\centering
\caption{Parameters from the fits to the eclipse light curves.}
\smallskip
\label{best_fit_table}
\begin{tabular}{llcccc}
\hline\hline
Line & Model & $F_{\rm line}$ & $r_{\rm inner}^\dagger$ & $r_{\rm outer}$ & 
     $\chi^2/{\rm dof}$ \\
     &       & ($10^{-13}$\,erg\,cm$^{-2}$s$^{-1}$) & ($10^{9}$\,cm) 
          & ($10^{11}$\,cm) & \\
\hline
\Heii & Flat & $0.71 \pm 0.02$ & 1 & $1.8\pm 0.2$ & $100/30$ \\
\Heii\ ($\phiorb > 0.92$) & Flat & $0.81 \pm 0.04$ & 1 & $1.6 \pm 0.3$ & $29/19$ \\
\Heii & Schandl-Meyer &  $0.57 \pm 0.02$ & 1 & $1.7^\dagger$ & $174/31$ \\
\Heii\ ($\phiorb > 0.92$) & Schandl-Meyer & $0.89 \pm 0.04$ & 1 & $1.7^\dagger$ & $50/20$ \\
\Heii & SLW & $0.90 \pm 0.02$  & 0.04 & $1.9\pm0.2$ & $109/30$ \\
\Heii\ ($\phiorb > 0.92$) & SLW & $0.92 \pm 0.04$ & 0.04 & $1.8\pm0.2$ & $39/19$\\
\Civ & Flat & $2.2 \pm 0.1$ & 1 & $1.6\pm <0.1$ & $627/30$ \\
\Nv & Flat & $3.8 \pm 0.1$ & 1 & $1.4\pm <0.1$  & $3290/30$  \\
\Nv & Flat, $\tau_c = 3$ & $3.8 \pm 0.1$ & 1 & $1.0\pm <0.1$ & $3270/30$ \\
\Siiv & Flat & $0.5 \pm 0.01$ & 1 & $1.6\pm <0.1$ & $386/30$ \\
\Siiv\ ($\phiorb > 0.92$) & Flat & $0.6 \pm 0.01$ & 1 & $1.4\pm <0.1$ & $197/19$ \\
\Ov & Flat & $0.4 \pm 0.01$ & 1 & $1.1\pm 0.1$ & $422/30$ \\
\Ov\ ($\phiorb > 0.92$) & Flat & $0.4 \pm 0.01$ & 1 & $1.1\pm 0.1$ & $191/19$ \\
\hline
\null$^\dagger$ fixed parameter
\end{tabular}
\end{table}

\begin{table}[p]
\centering
\caption{$\chi^2$ statistics for the various line profiles.}
\smallskip
\label{chi2_values}
\begin{tabular}{llccccc}
\hline\hline
Line & Model & $(v_{\rm min}, v_{\rm max})^\ddagger$ & \multicolumn{4}{c}{$\chi^2/{\rm dof}$} \\
     &       & $10^3$\,km\,s$^{-1}$ & $\phiorb = 0.909$ & 0.939 & 0.057 & 0.092 \\
\hline
\Heii & Flat & ($-1.5,1.5$) & 111/27 & 32/27 & 24/45 & 75/45 \\
\Heii ($\phiorb > 0.92$) & Flat & ($-1.5,1.5$) & 127/27 & 31/27 & 26/45 & 71/45 \\ 
\Heii & Schandl-Meyer & ($-1.5,1.5$) & 153/27 & 47/27 & 17/45 & 68/45 \\
\Heii ($\phiorb > 0.92$) & Schandl-Meyer & ($-1.5,1.5$) & 306/27 & 33/27 & 41/45 & 83/45 \\ 
\Heii  & SLW & ($-1.5,1.5$) & 102/27 & 33/27 & 43/45 & 155/45 \\
\Heii ($\phiorb > 0.92$) & SLW & ($-1.5,1.5$) & 102/27 & 32/27 & 45/45 & 163/45 \\
\Civ  & Flat & ($-2.0,2.0$) & 491/35 & 223/35 & 404/64 & 243/64 \\ 
\Nv   & Flat & ($-1.5, 3.0$) & 2831/31 & 1505/31 & 1637/70 & 2251/70 \\ 
\Nv   & Flat, $\tau_c = 3$ & ($-1.5,3.0$) & 50537/31 & 10676/31 & 1022/70 & 30918/70 \\ 
\Siiv & Flat & ($-1.5,0.8$) & 423/17 & 94/17 & 135/35 & 243/35 \\ 
\Siiv ($\phiorb > 0.92$) & Flat & ($-1.5,0.8$) & 559/17 & 74/17 & 132/35 & 247/35 \\ 
\Ov   & Flat & ($-1.5,1.5$) & 316/23 & 165/23 & 51/46 & 457/46 \\ 
\Ov   ($\phiorb > 0.92$) & Flat & ($-1.5,1.5$) & 342/23 & 187/23 & 55/46 & 449/46 \\ 
\hline
\multicolumn{3}{l}\null$^\ddagger$ velocity range over which $\chi^2$ was computed\\
\end{tabular}
\end{table}

\clearpage

\figcaption[fig1.ps]{(a) Eclipse light curve of the
\Heii\,$\lambda$\,1640.47 emission line data and model fits.  The
solid curve is the model light curve fit to data for $\phiorb > 0.92$
(i.e., excluding the ``dip''), and the dotted curve is the model fit
to all of the data.  (b) \Civ\,$\lambda\lambda$\,1548.195,1550.770
eclipse data and model light curves.
(c)\Nv\,$\lambda\lambda$\,1238.821,1242.804 eclipse light curves.  The
solid curve is for a marginally optically thick emission line (see
text) while the dotted curve is for an optically thick emission line.
Although both light curves fit the data comparably well, the best-fit
outer disk radii are significantly different: $r_{\rm outer} = 1.0
\times 10^{11}$\,cm for the marginally optically thick line, while
$r_{\rm outer} = 1.4\times 10^{11}\,$cm for the optically thick line.
(d) \Siiv\,$\lambda$\,1393.755 eclipse light curves.  As with \Heii,
the solid curve is fit to data for $\phiorb > 0.92$ and the dotted
curve is the model fit to all of the data.  (e)
\Ov\,$\lambda$\,1371.292 eclipse light curves.  The solid and dotted
curves are as for the \Heii\ and \Siiv\ lines.  (f) UV continuum
eclipse light curves obtained from deriving the fluxes for the \Heii\
(crosses), \Civ\ (stars), \Nv\ (diamonds), \Siiv\ (triangles), and
\Ov\ (squares) emission lines.  The gradual decline and rise in
continuum flux during ingress and egress imply that the UV emission is
extended over a disk region comparable in size to the emission line
regions.
\label{eclipse_lcs}}

\figcaption[fig2.ps]{Model versus \Heii\ emission line data at various
eclipse phases.  Left: The images show the apparent disk surface
brightness taking into account the effects of a disk wind.  Lighter
regions indicate strong emission while dark regions indicate
suppressed emission.  The ``eye'' symbol shows the orientation of the
observer at each phase relative to the line-of-centers which extends
horizontally to the right from the disk center.  A line passing
through the center of the disk and the ``eye'' corresponds to the disk
locations with zero projected velocity as seen by the observer
(neglecting the Doppler shifts induced by the orbital motion).  The
orientation of the eye relative to the surface brightness pattern can
be understood from the zeroes of Eq.~\ref{disk_wind_shear} as a
function of the azimuthal angle $\phi$ which is measured with respect
to this line.  The dark bands along two of the disk diameters are
regions of highly suppressed emission and are given by $\phi = 0$ and
$\phi = \tan^{-1}[(-\partial v_r/\partial r)/(3v_\phi/2r)]$; the
observer orientation is perpendicular to the $\phi = 0$ band.  The
disk regions which are eclipsed by the companion are shown in black.
Center: \Heii\ emission lines as a function of projected velocity.
Right: The residual spectra after subtracting the model line emission.
For reference, the dotted vertical lines are the Doppler shifts of the
neutron star due to the orbital motion of the system; these velocities
should correspond to the line center for emission from the disk.
\label{HeII_profiles}}

\figcaption[fig3.ps]{Model versus \Civ\ emission line profiles.  These 
calculations include interstellar absorption obtained from fitting the
out-of-eclipse profile at $\phiorb = 0.132$ (see text).  The velocity
scale is referred to the rest wavelength of the stronger doublet
member, 1548.195\AA.
\label{Civ_profiles}}

\figcaption[fig4.ps]{Mean residuals after subtraction of the model line 
profiles for the ingress phases $\phiorb = 0.902$--0.916.  The dashed
lines are the estimates for the respective continua.  Left: \Heii\
emission line. Right: \Civ\ line.  These residuals can be interpreted
as blue-shifted absorption and excess red emission due to scattering
in a disk wind.
\label{early_phase_residuals}}

\figcaption[fig5.ps]{Disk wind line profiles for optically thick 
($\tau_c = 10^6$, dashed curve), marginally thick ($\tau_c = 3$,
solid), and optically thin ($\tau_c = 10^{-6}$, dotted) emission.
\label{tau_profiles}}

\figcaption[fig6.ps]{Model versus \Nv\ emission line profiles.  The
disk surface brightness images shown on the left are for a disk model
with $\tau_c = 3$ (Eq.~\ref{tau_scaling}) and therefore differ from
the optically thick images shown in Fig.~\ref{HeII_profiles}.  The
solid curves in the middle and right panels are the marginally
optically thick model with $\tau_c = 3$, while the dashed curves are
for the optically thick model.  For the marginally optically thick
model, a doublet ratio of 1.22 is used.  The error bars on the data
are suppressed for clarity.  The mean error bar size is shown in the
upper right corner of each plot.  The velocity scale is referred to
the rest wavelength of the stronger doublet member, 1238.821\AA.
\label{Nv_profiles}}

\figcaption[fig7.ps]{Eclipse light curves for the (a) Schandl-Meyer and 
(b) SLW warped disk models fit to the \Heii\ fluxes.  The outer radius
in the Schandl-Meyer models is fixed at $1.7 \times 10^{11}$\,cm.  The
solid and dashed curves correspond to fits to two subsets of the data
used in order to find the out-of-eclipse flux normalization.  Both the
flux and outer radius are allowed to vary for the SLW fits to these
data, yielding nearly identical results for both subsets of the data.
\label{warp_light_curves}}

\figcaption[fig8.ps]{Apparent disk surface brightness, line profiles and
residuals for the \Heii\ line in the Schandl-Meyer warped disk model
(cf. Fig.~\ref{HeII_profiles}). 
\label{SM_warp_line_profiles}}

\figcaption[fig9.ps]{Apparent disk surface brightness, line profiles and 
residuals for the \Heii\ line in the SLW warped disk model.
(cf. Fig.~\ref{HeII_profiles}).
\label{SLW_warp_line_profiles}}

\setcounter{figure}{0}

\begin{figure}
\centerline{\epsfxsize=2.75in\epsfbox{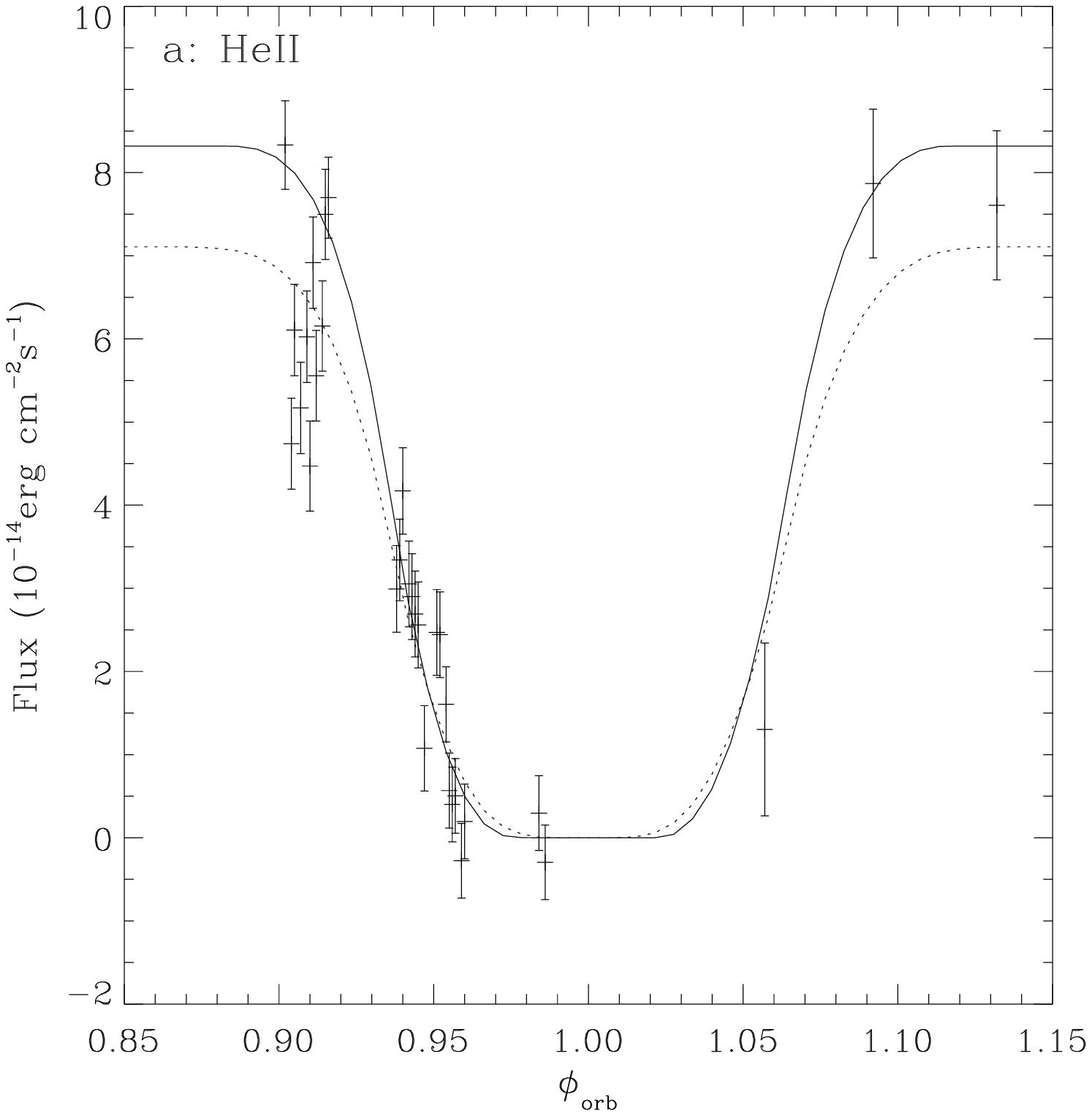}
            \epsfxsize=2.75in\epsfbox{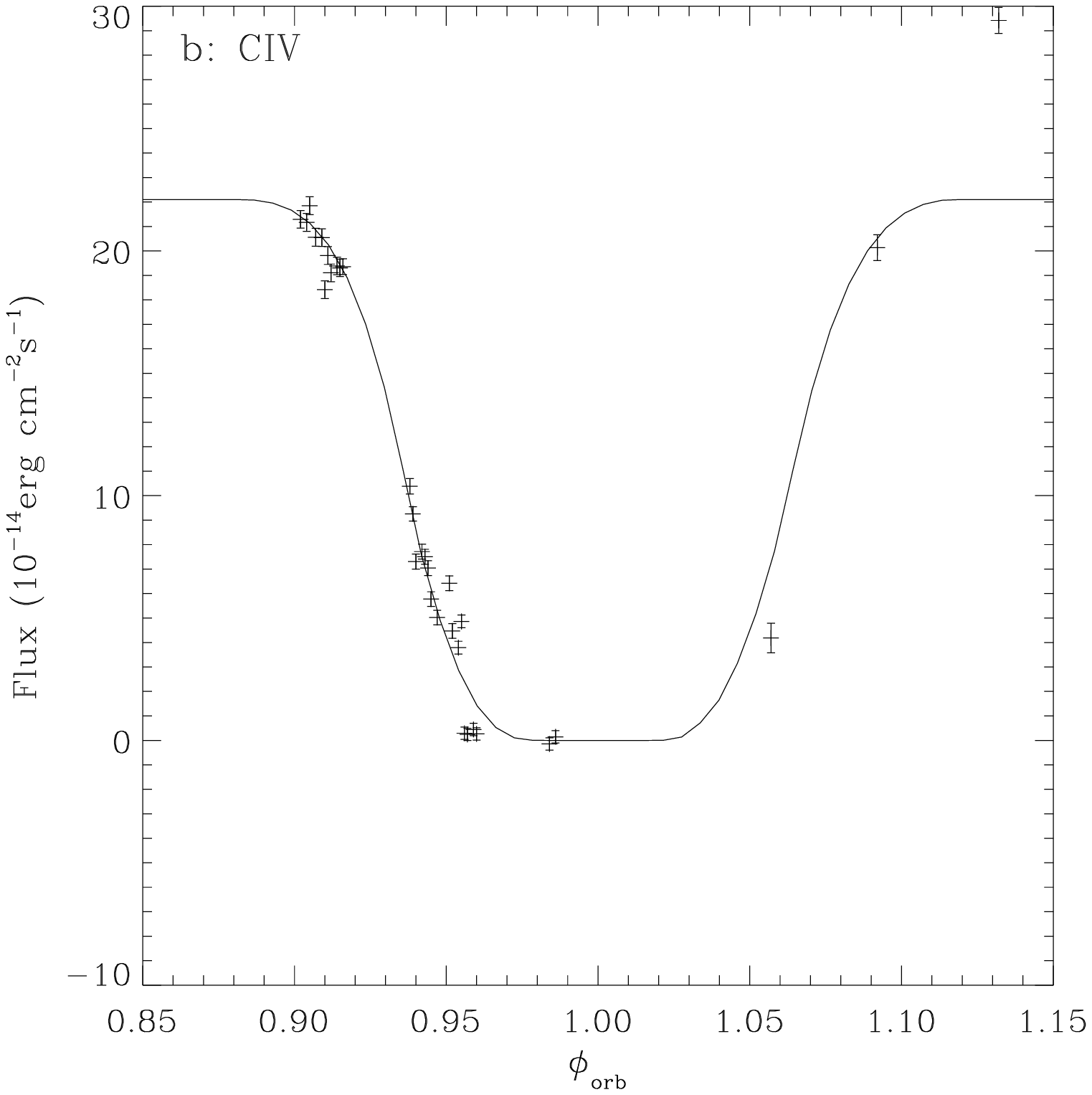}}
\centerline{\epsfxsize=2.75in\epsfbox{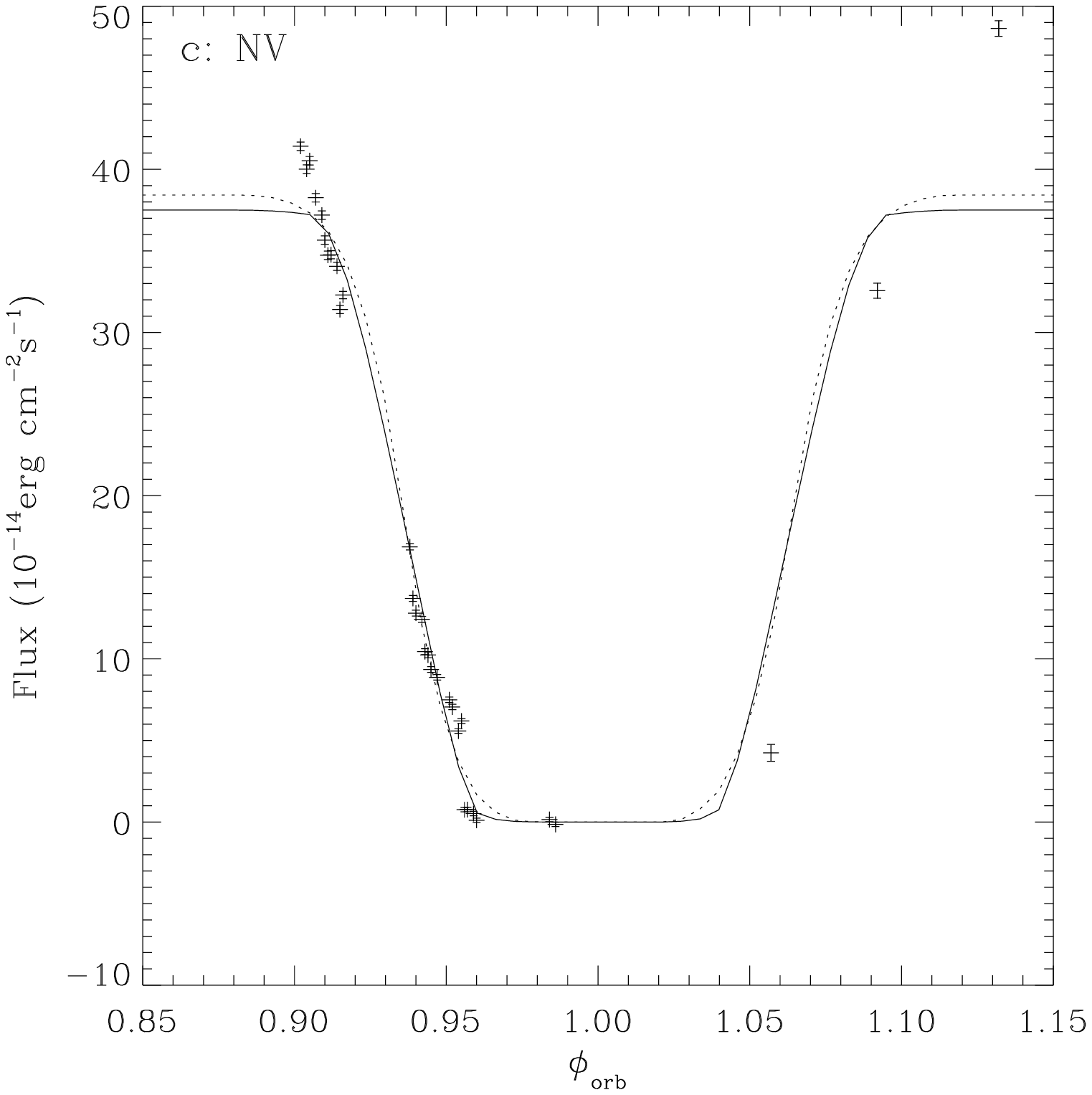}
            \epsfxsize=2.75in\epsfbox{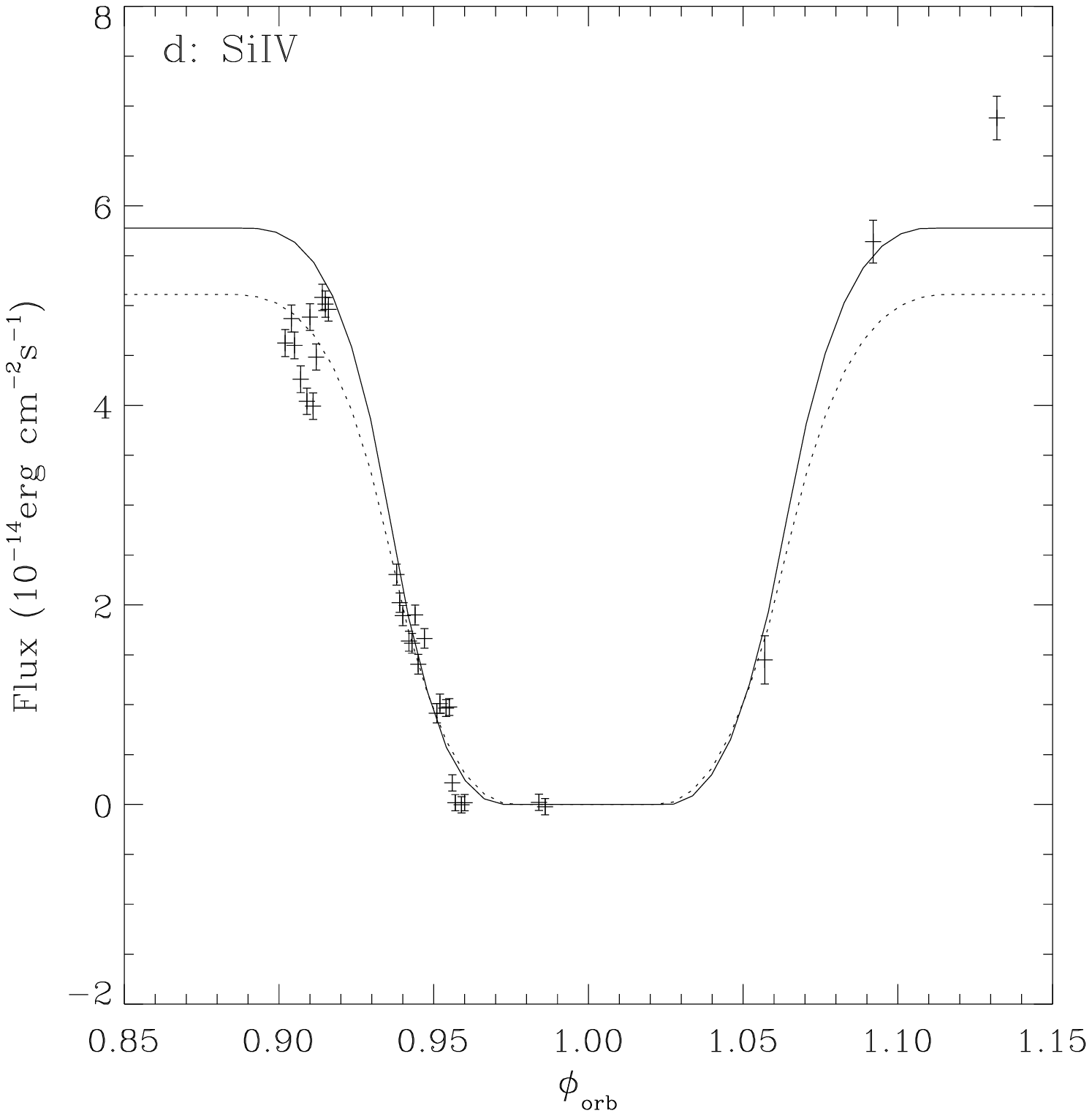}}
\centerline{\epsfxsize=2.75in\epsfbox{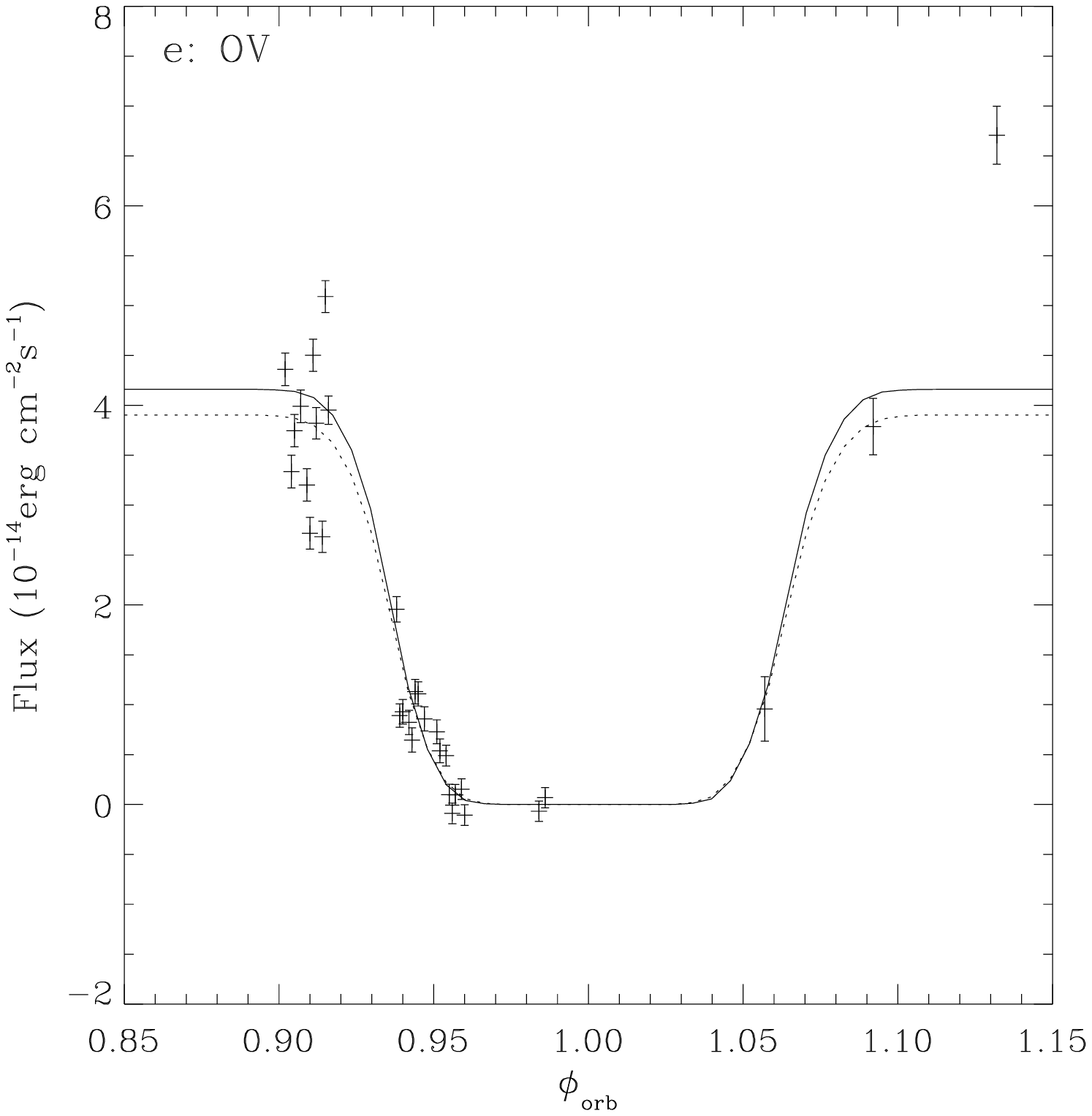}
            \epsfxsize=2.75in\epsfbox{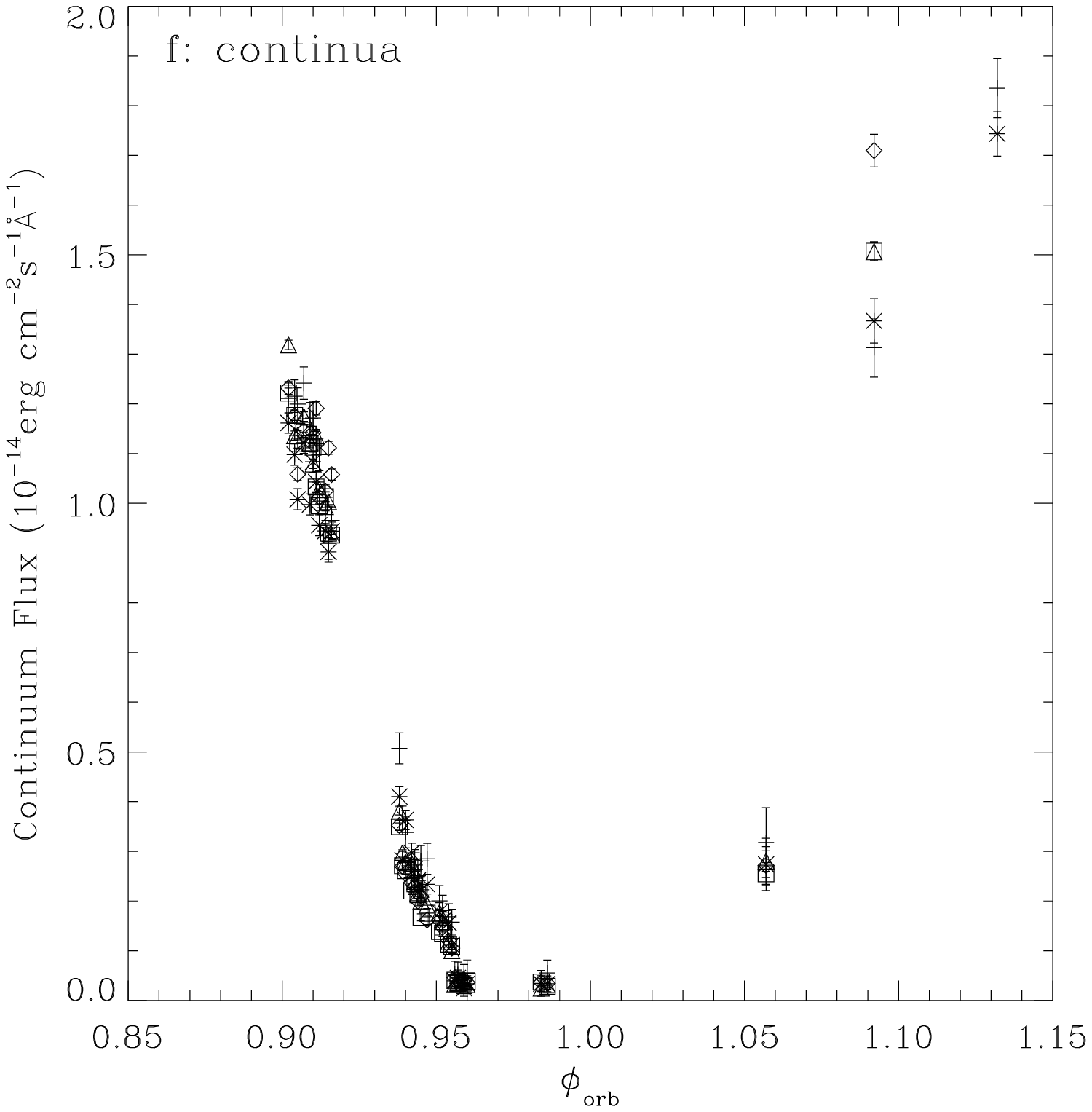}}
\caption{}
\end{figure}

\begin{figure}
\centerline{\epsfxsize=6in\epsfbox{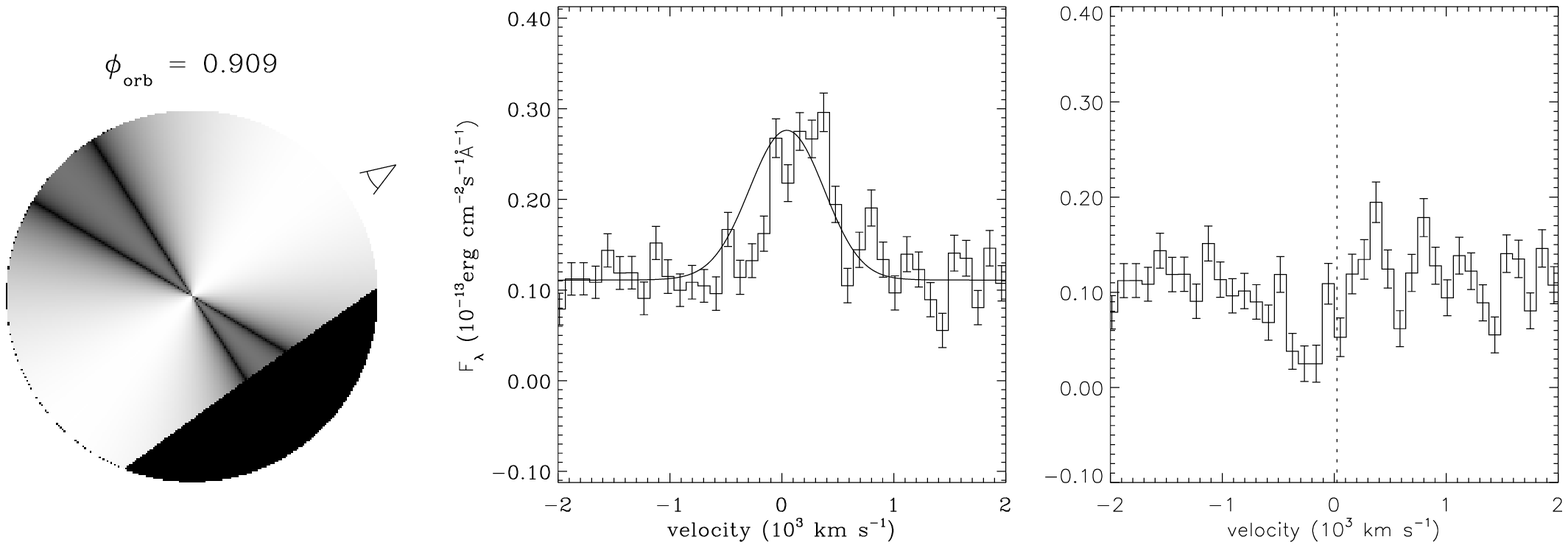}}
\centerline{\epsfxsize=6in\epsfbox{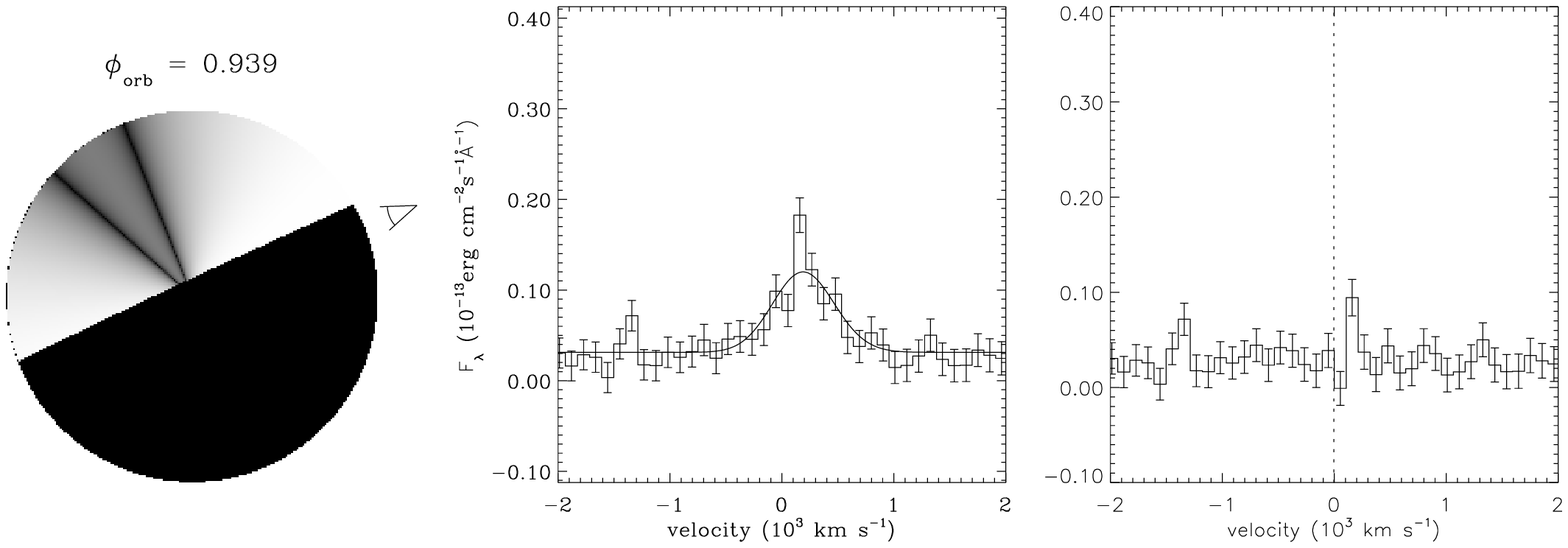}}
\centerline{\epsfxsize=6in\epsfbox{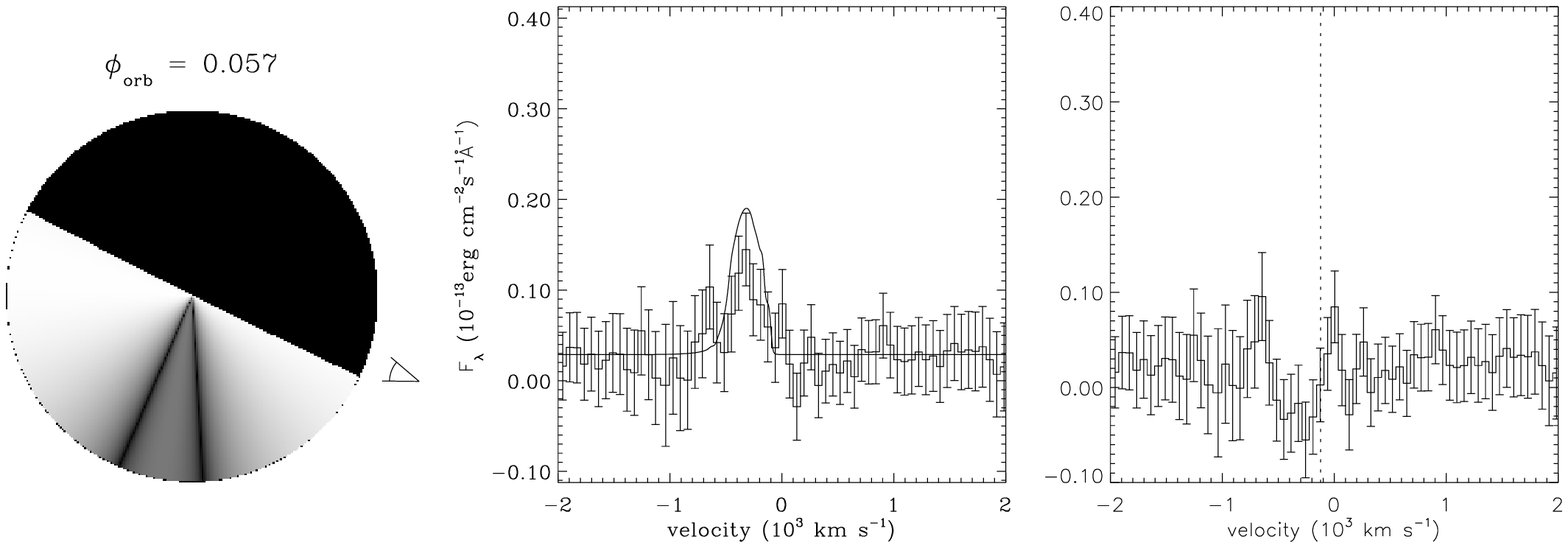}}
\centerline{\epsfxsize=6in\epsfbox{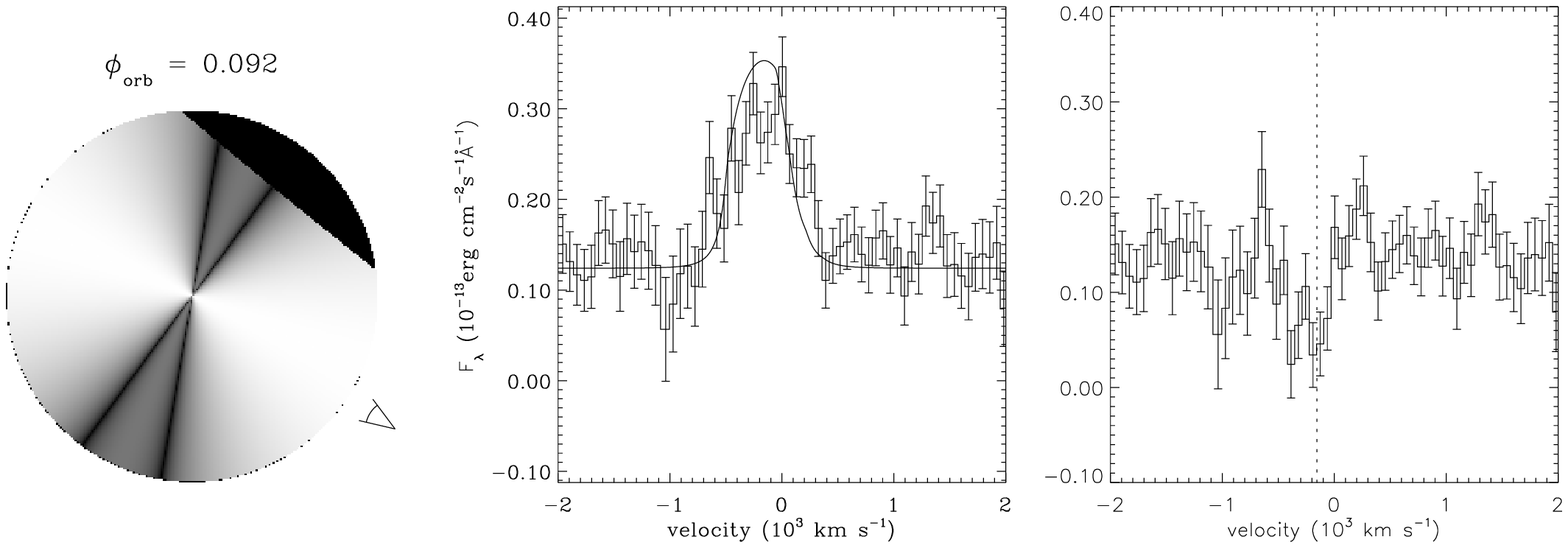}}
\caption{}
\end{figure}

\begin{figure}
\centerline{\epsfxsize=3in\epsfbox{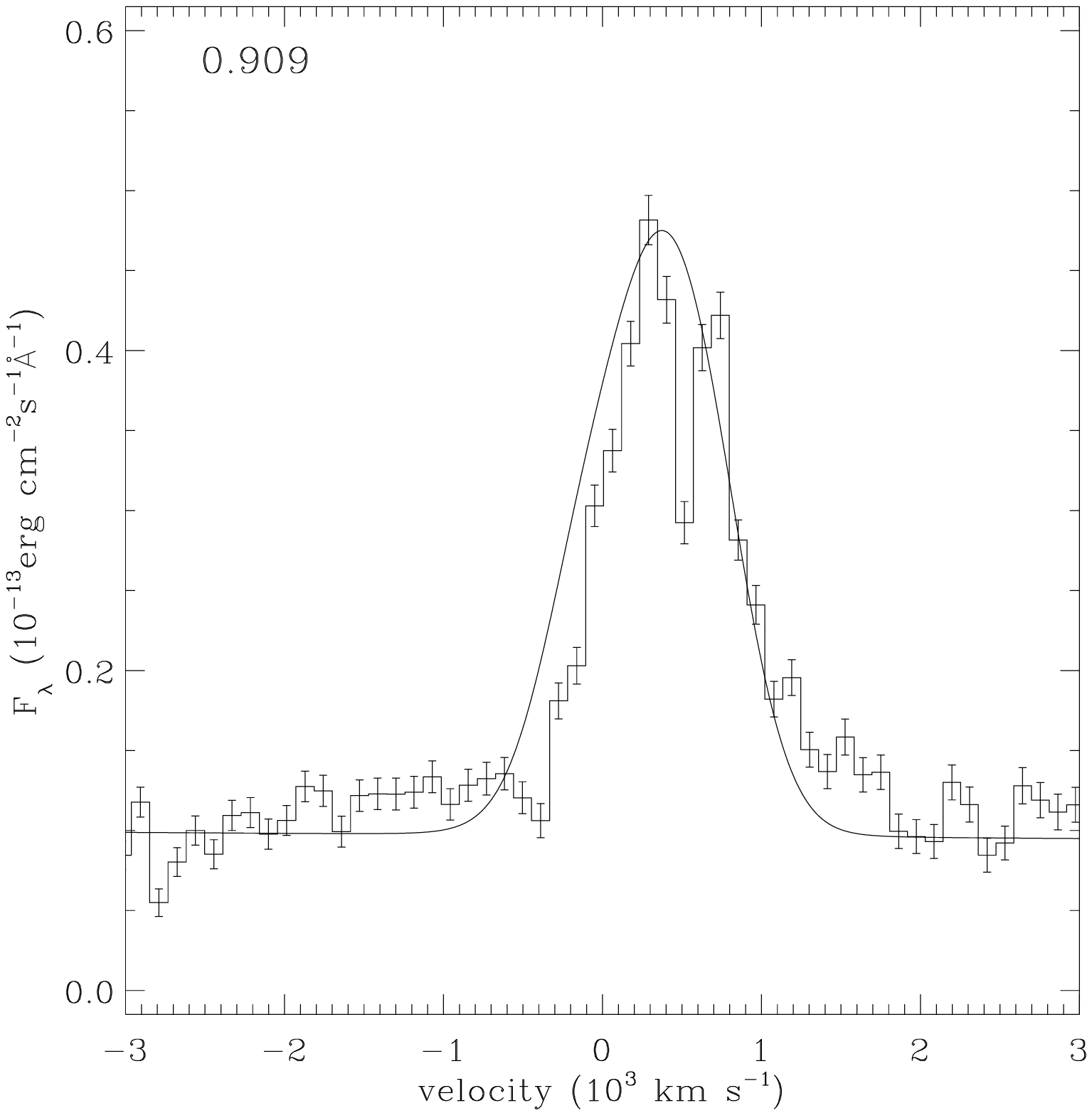}
            \epsfxsize=3in\epsfbox{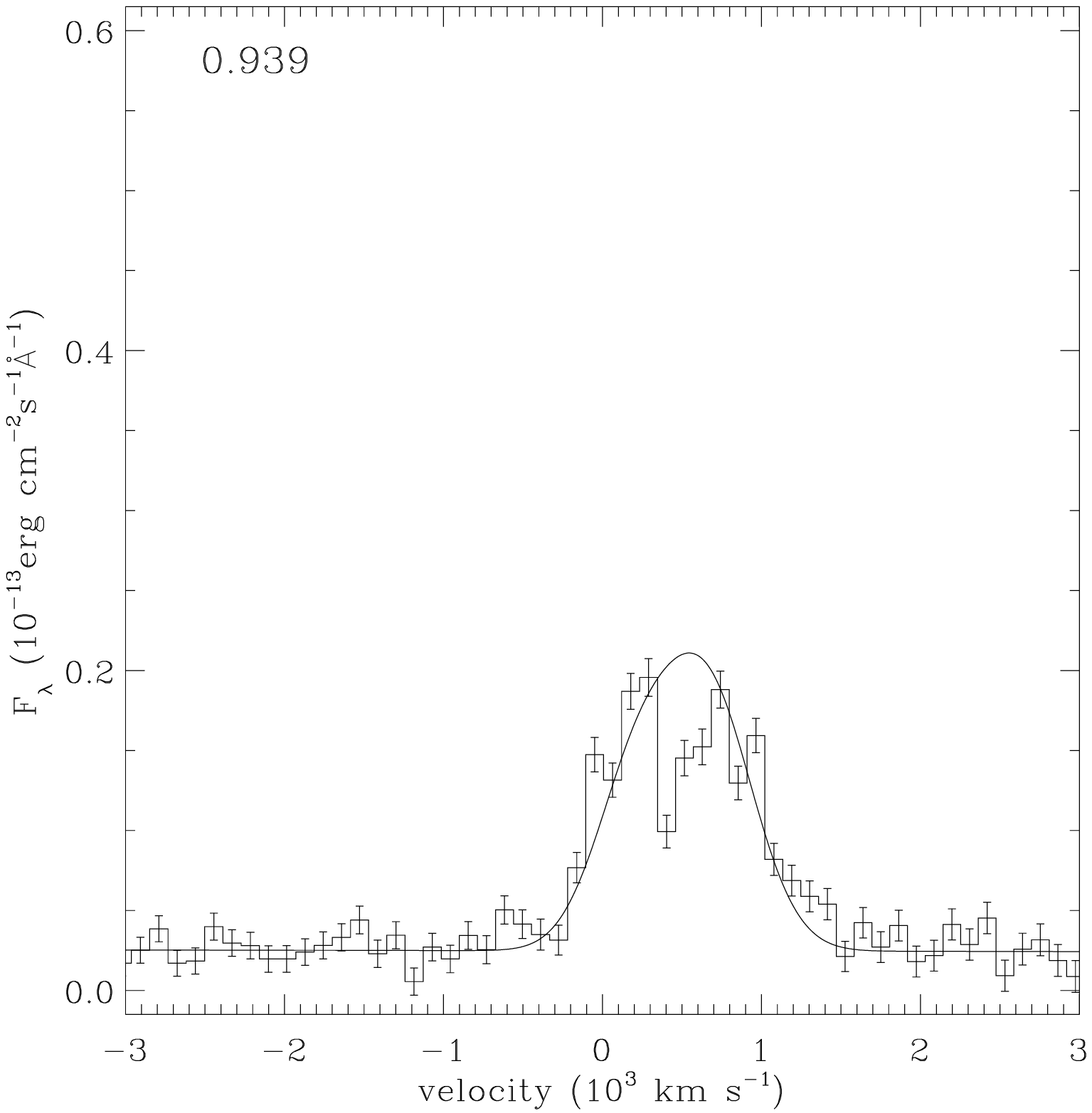}}
\centerline{\epsfxsize=3in\epsfbox{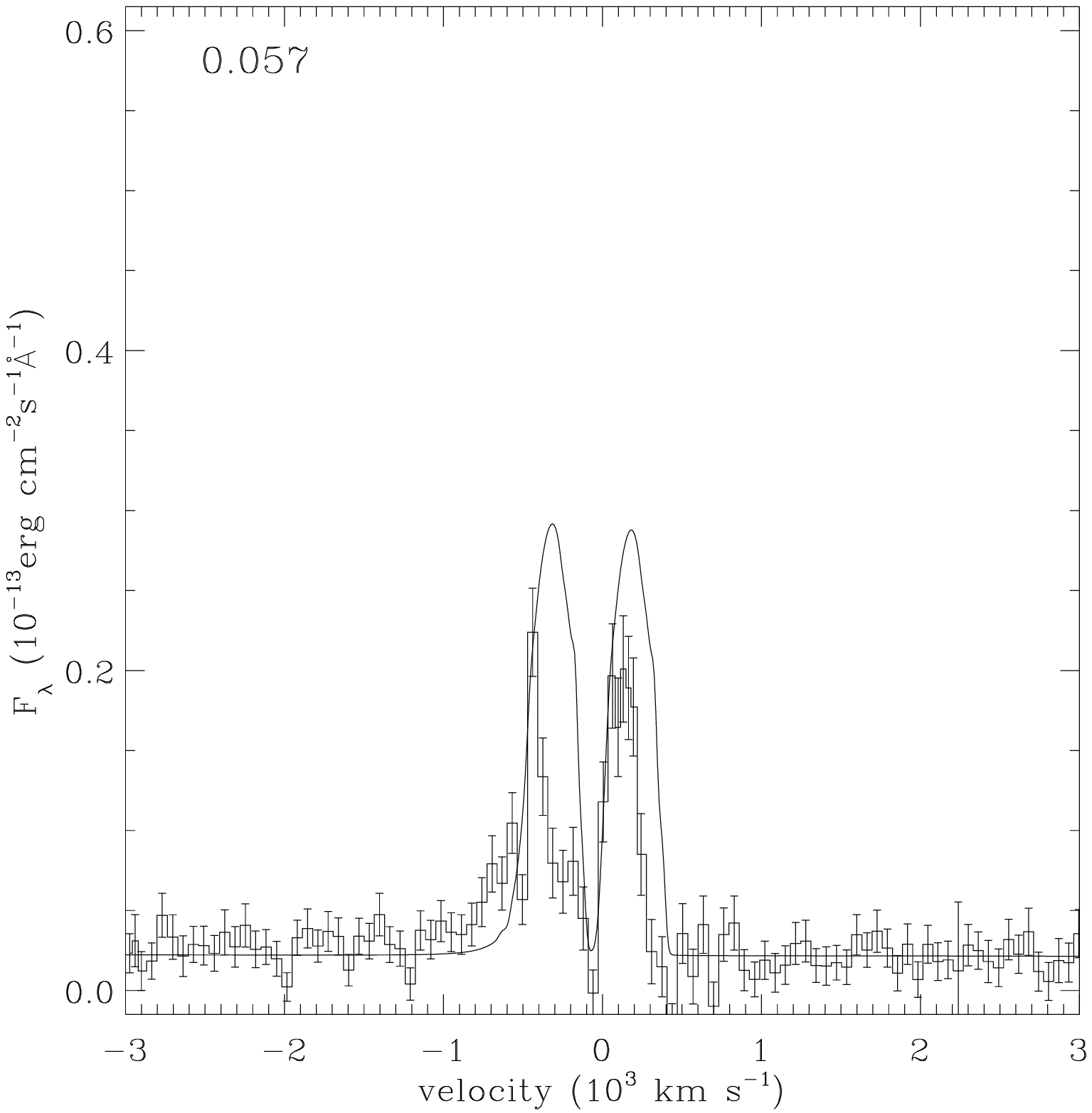}
            \epsfxsize=3in\epsfbox{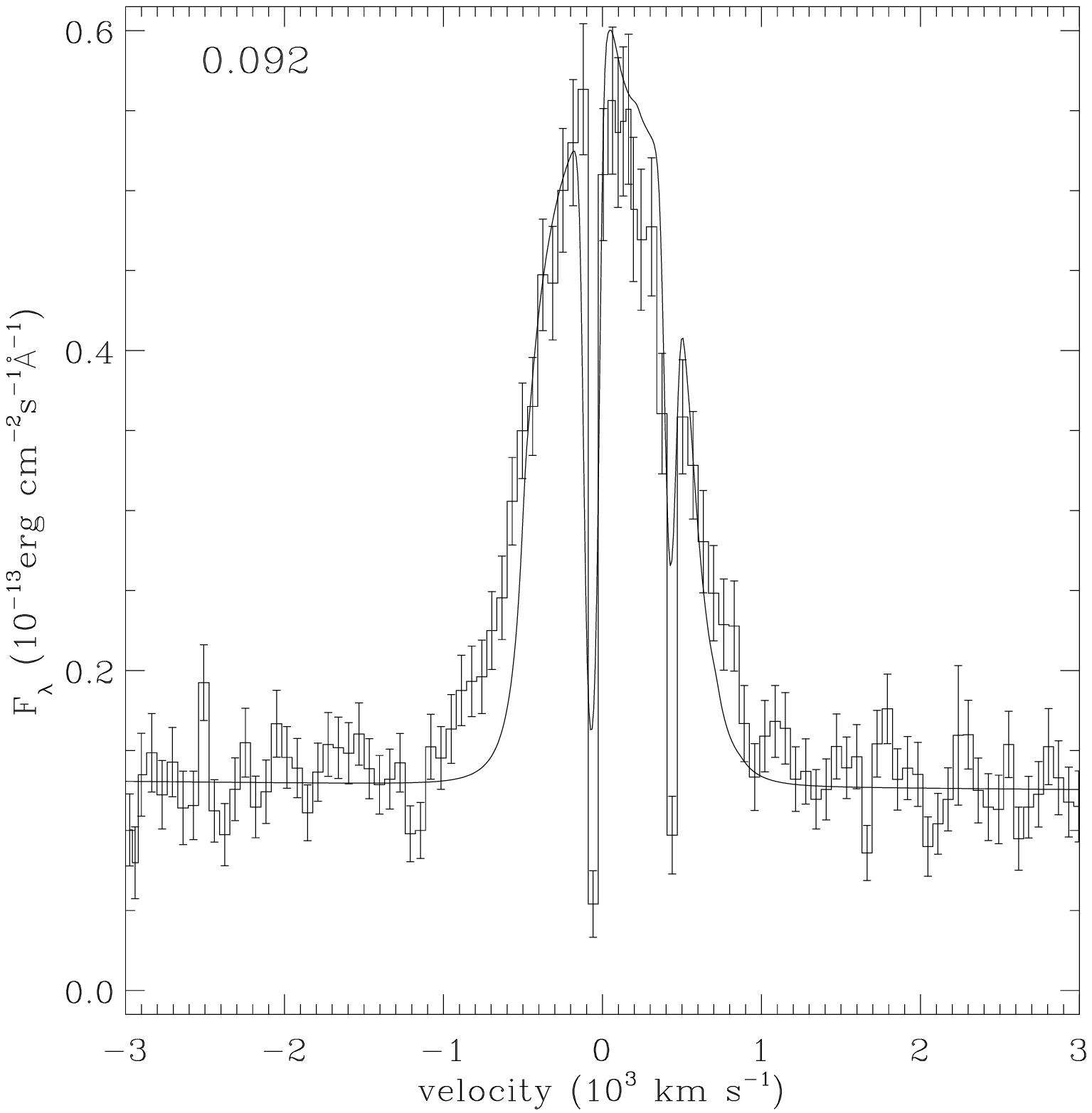}}
\caption{}
\end{figure}

\begin{figure}
\centerline{\epsfxsize=3in\epsfbox{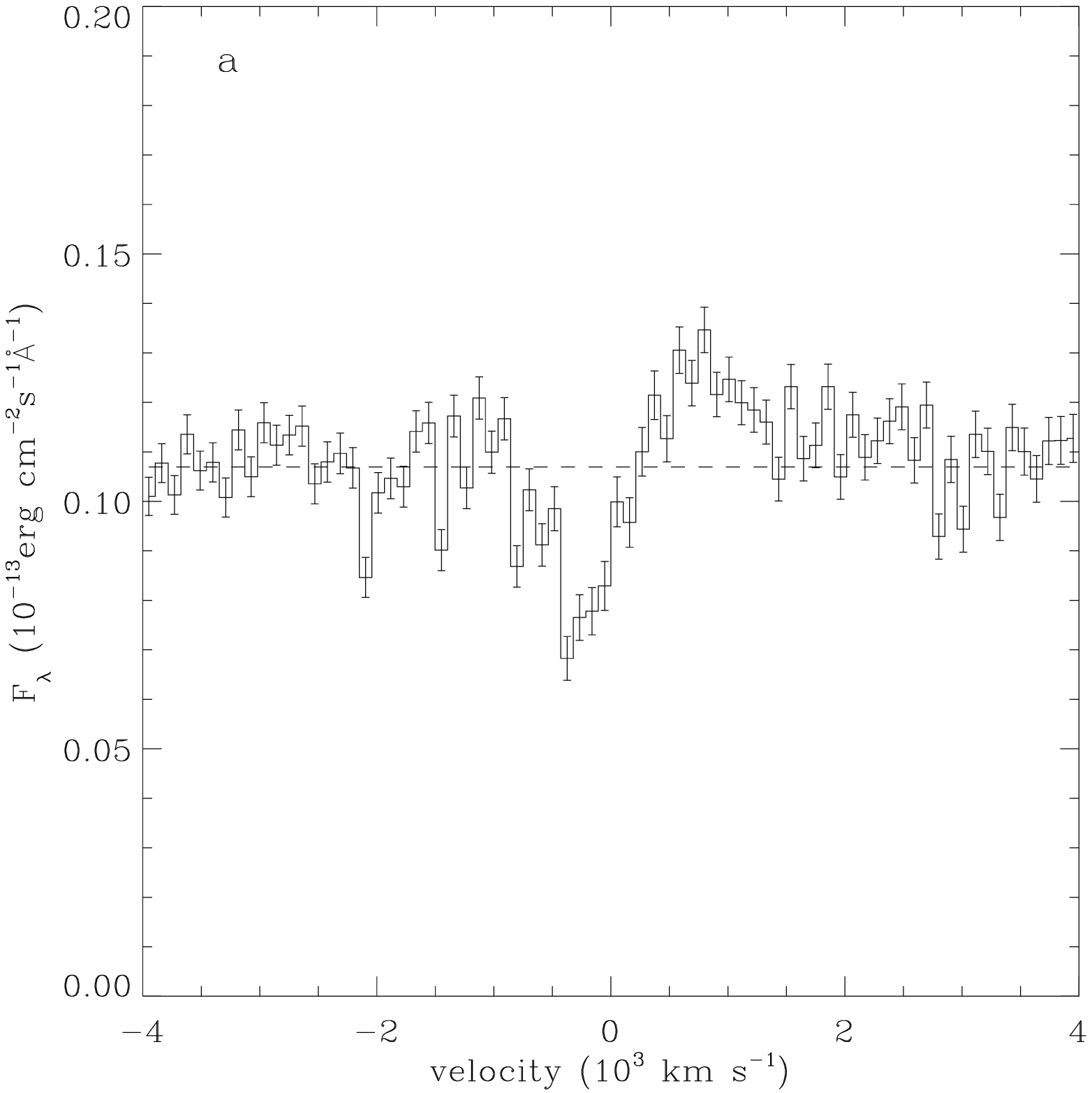}
            \epsfxsize=3in\epsfbox{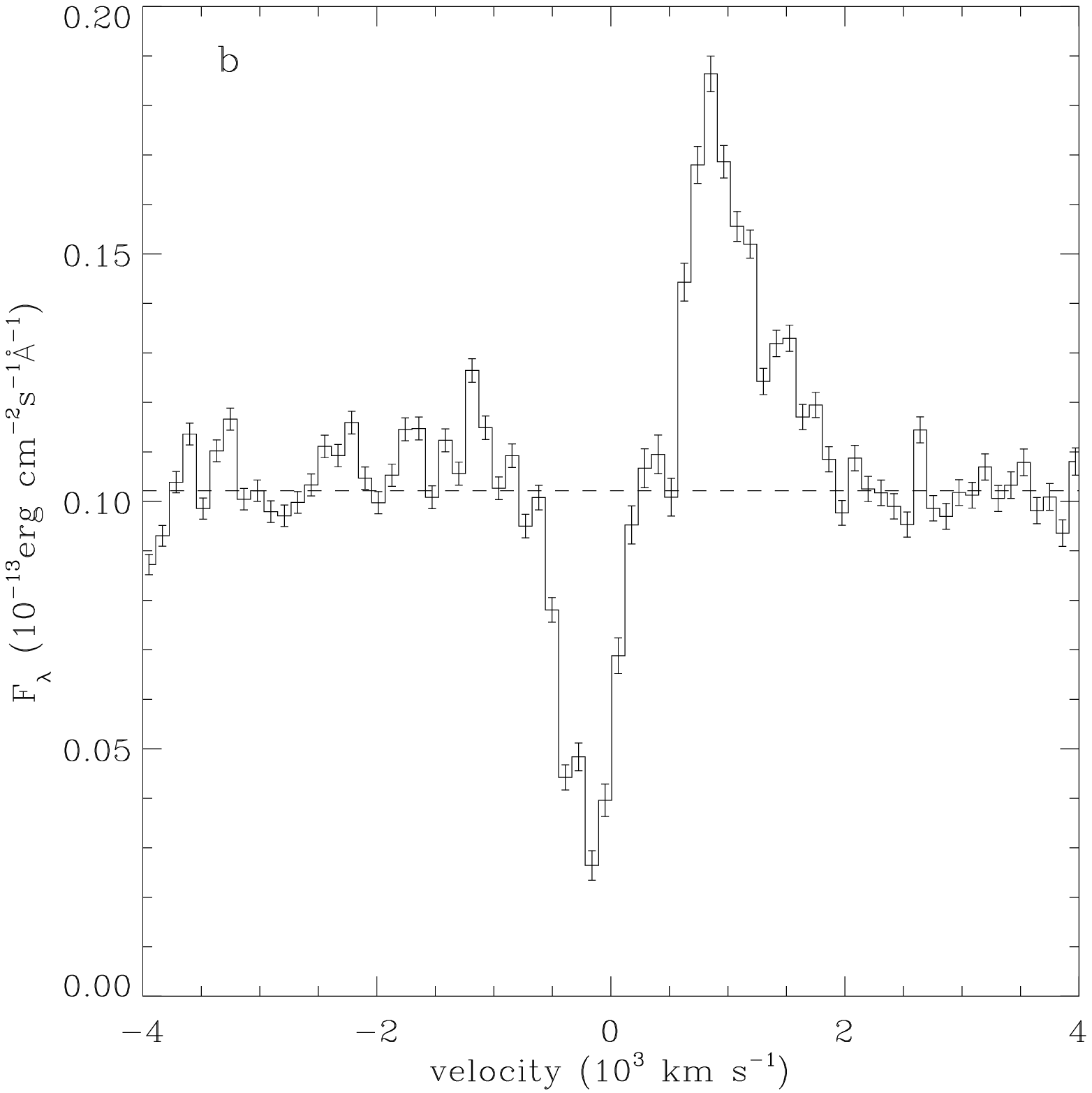}}
\caption{}
\end{figure}

\begin{figure}
\centerline{\epsfxsize=4in\epsfbox{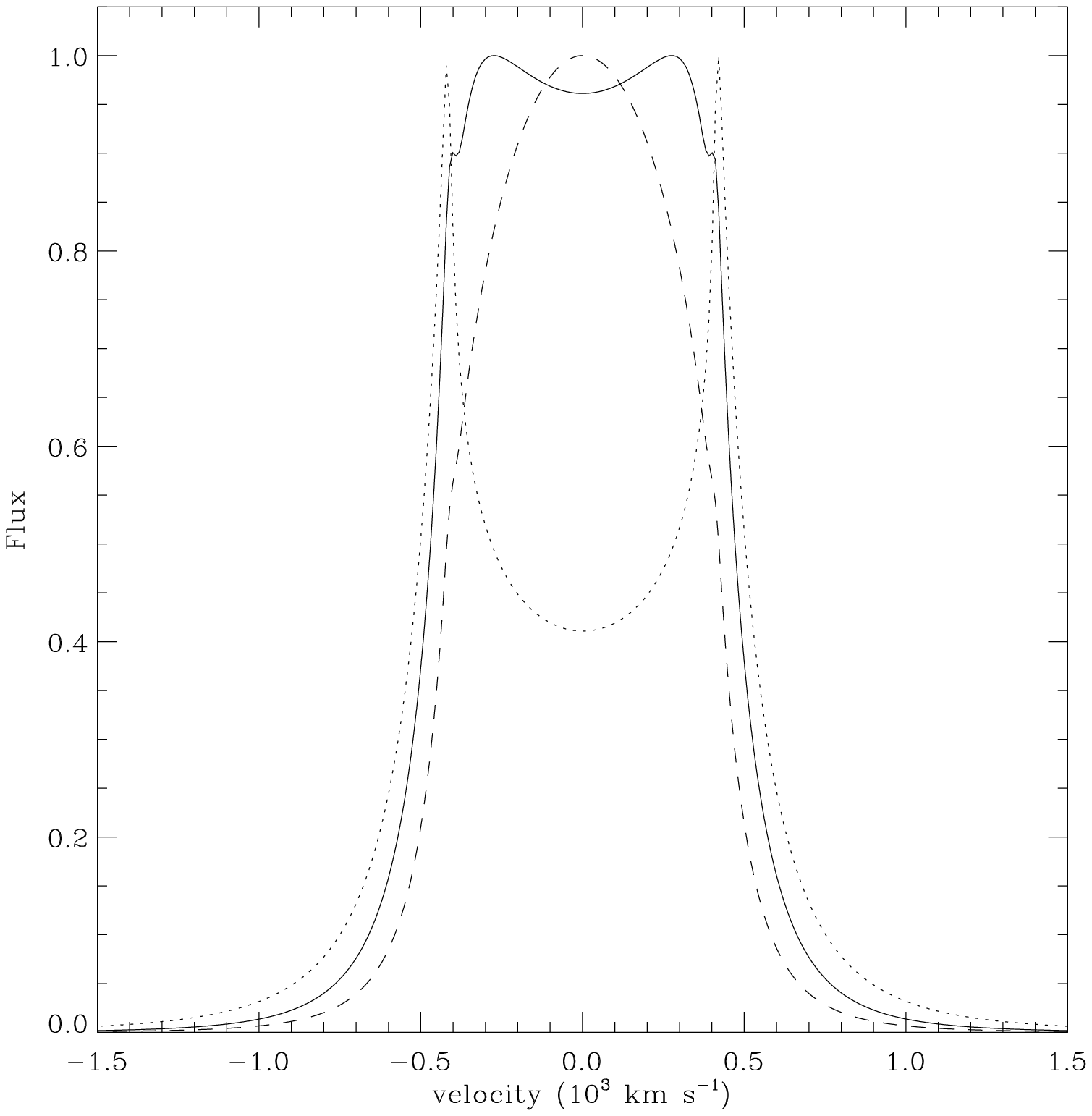}}
\caption{}
\end{figure}

\begin{figure}
\centerline{\epsfxsize=6in\epsfbox{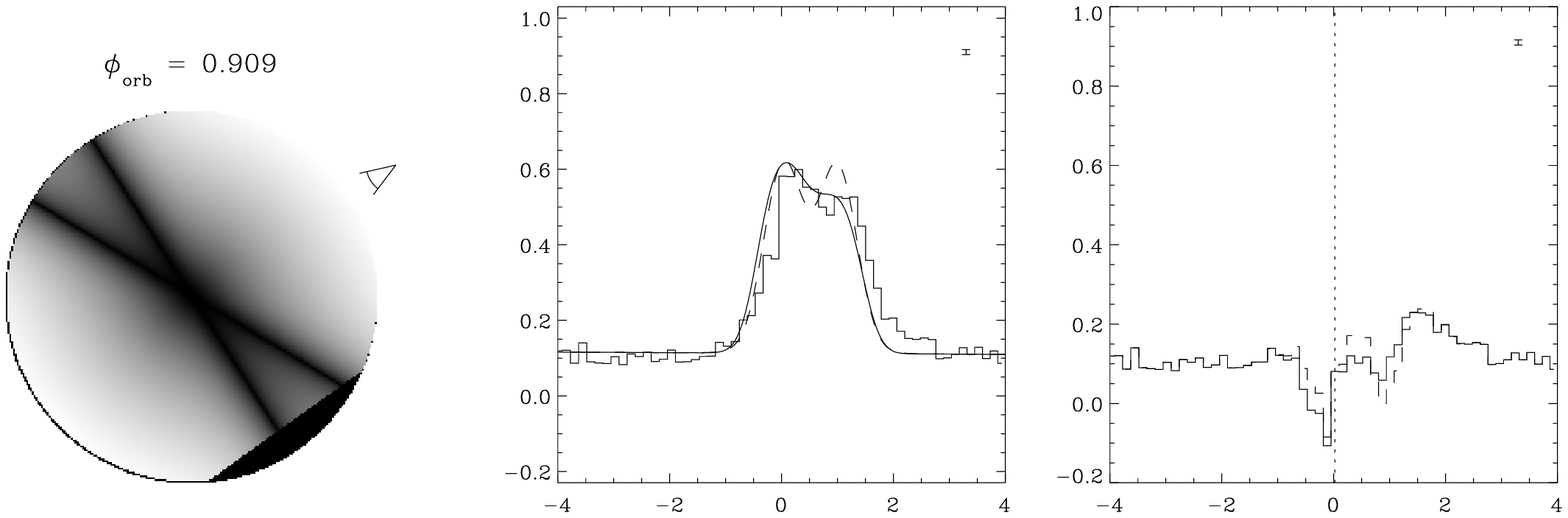}}
\centerline{\epsfxsize=6in\epsfbox{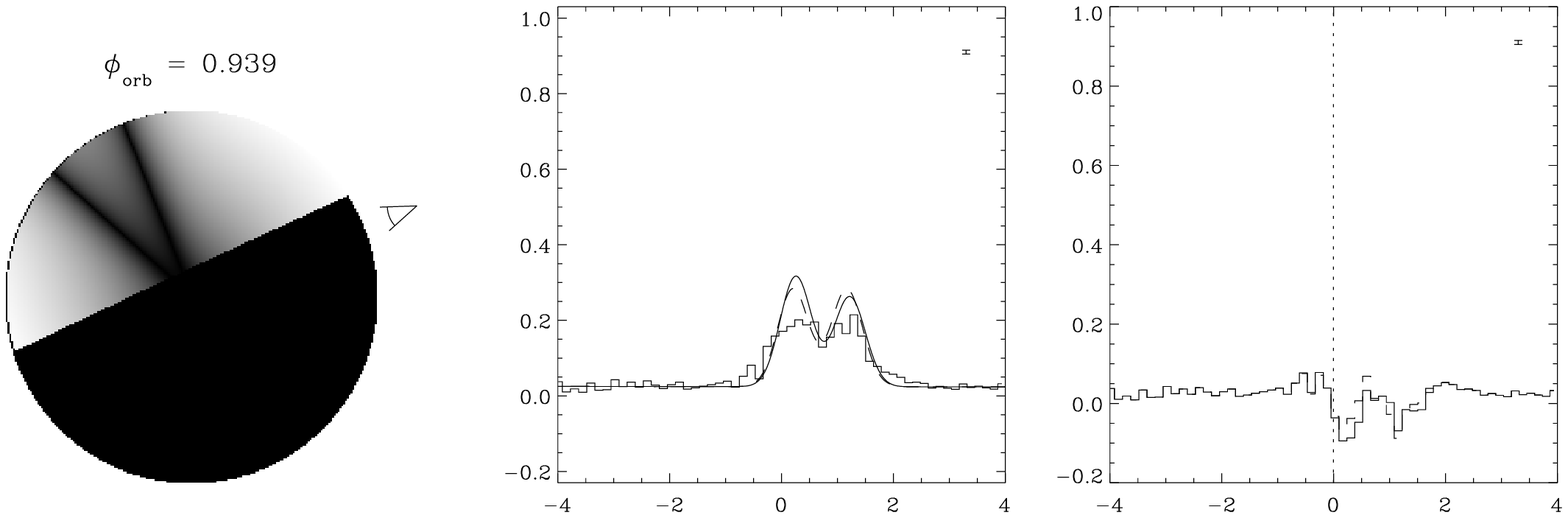}}
\centerline{\epsfxsize=6in\epsfbox{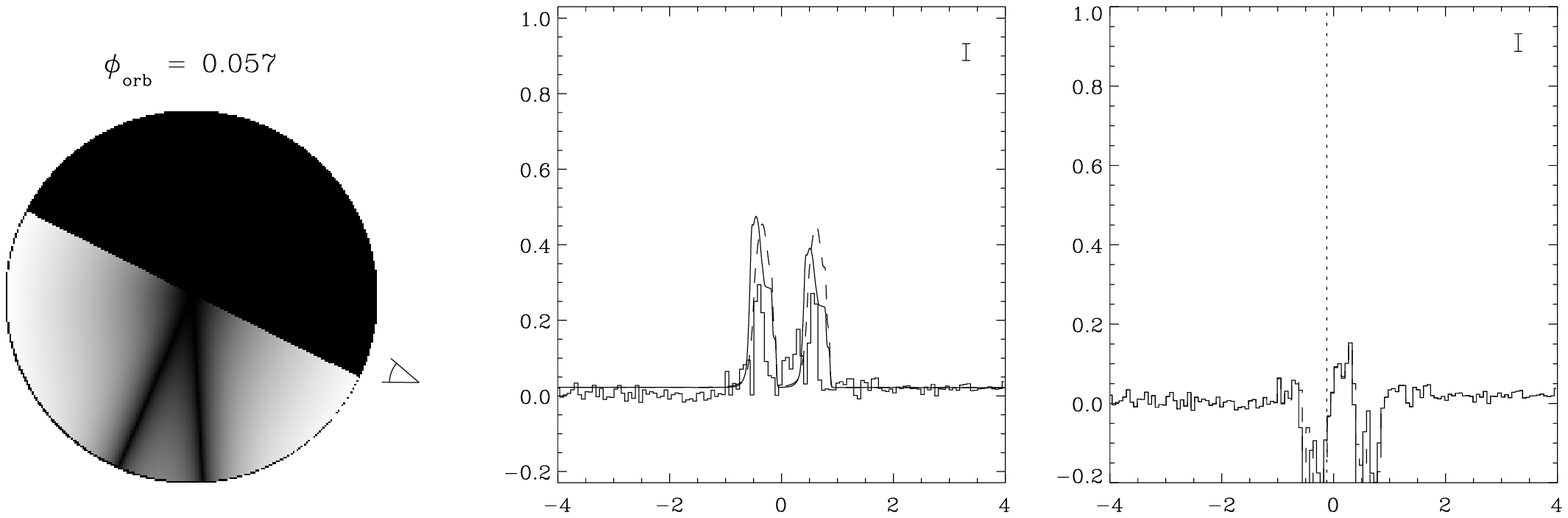}}
\centerline{\epsfxsize=6in\epsfbox{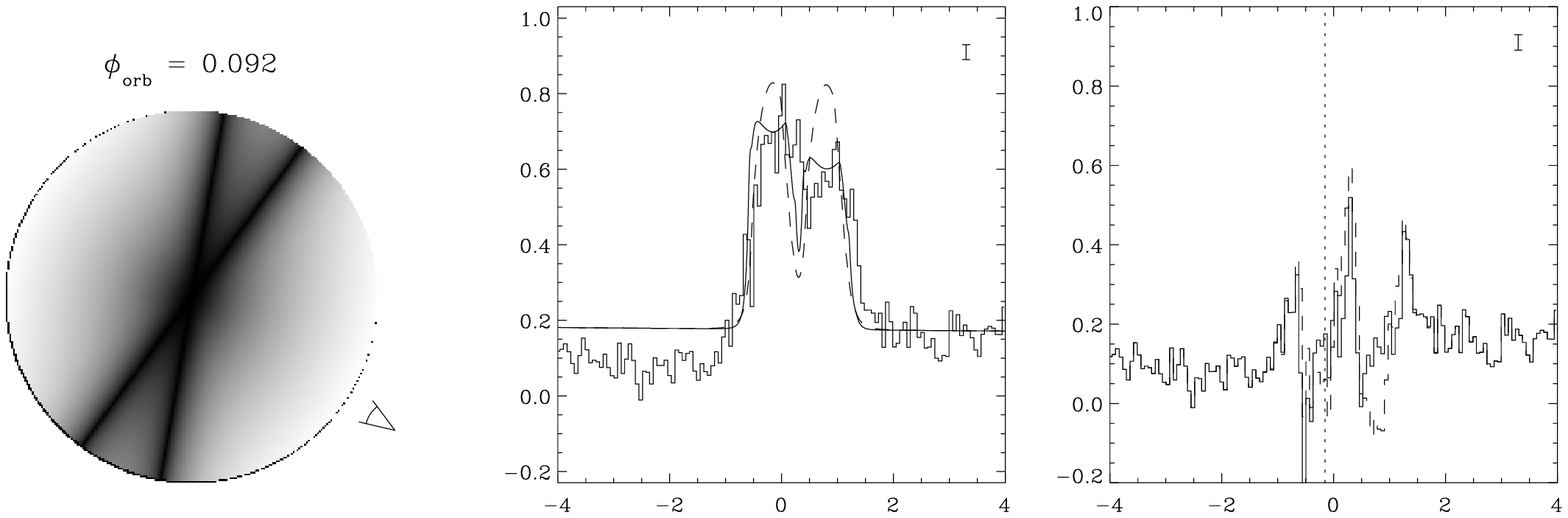}}
\caption{}
\end{figure}

\begin{figure}
\centerline{\epsfxsize=3in\epsfbox{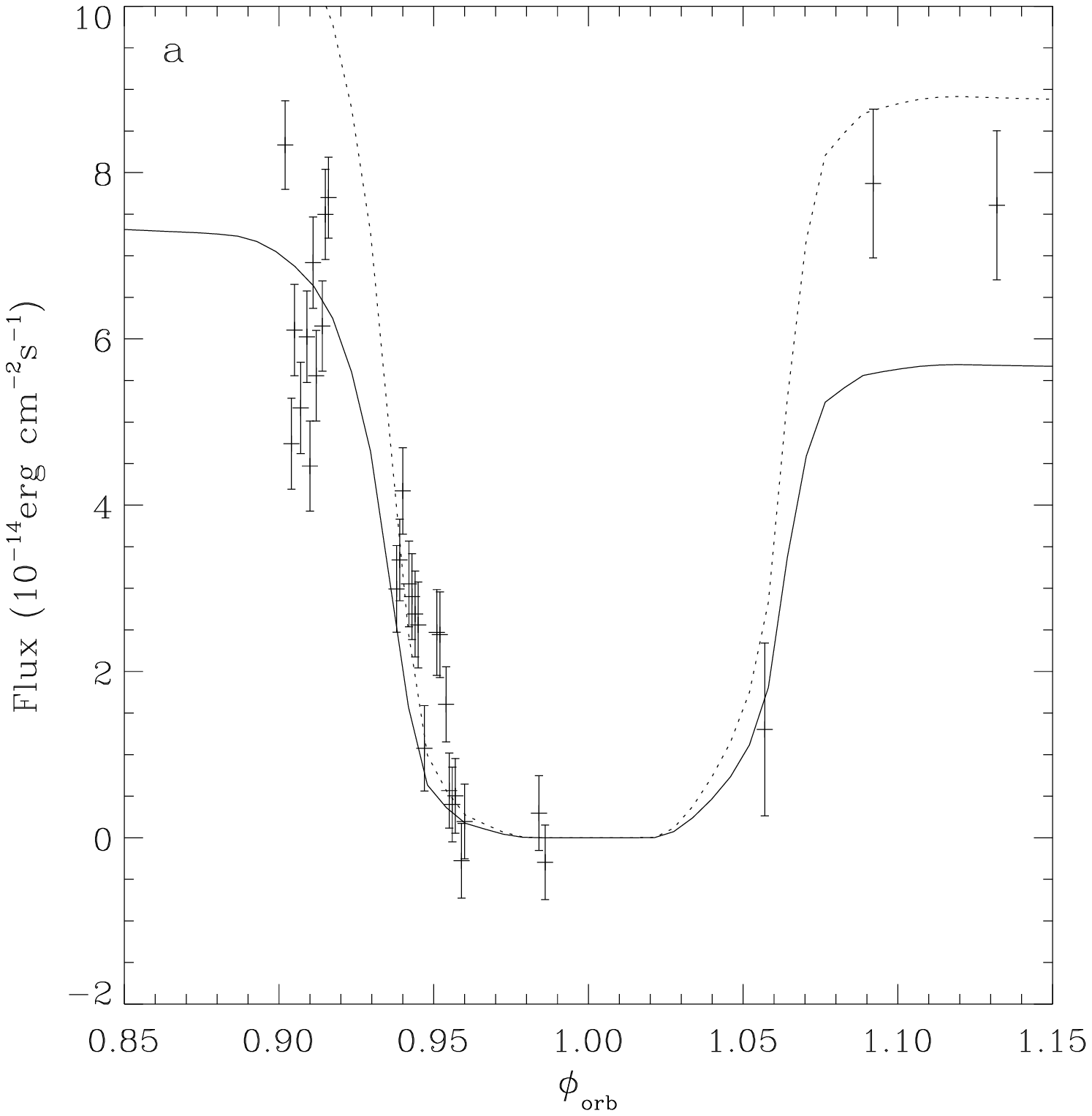}
            \epsfxsize=3in\epsfbox{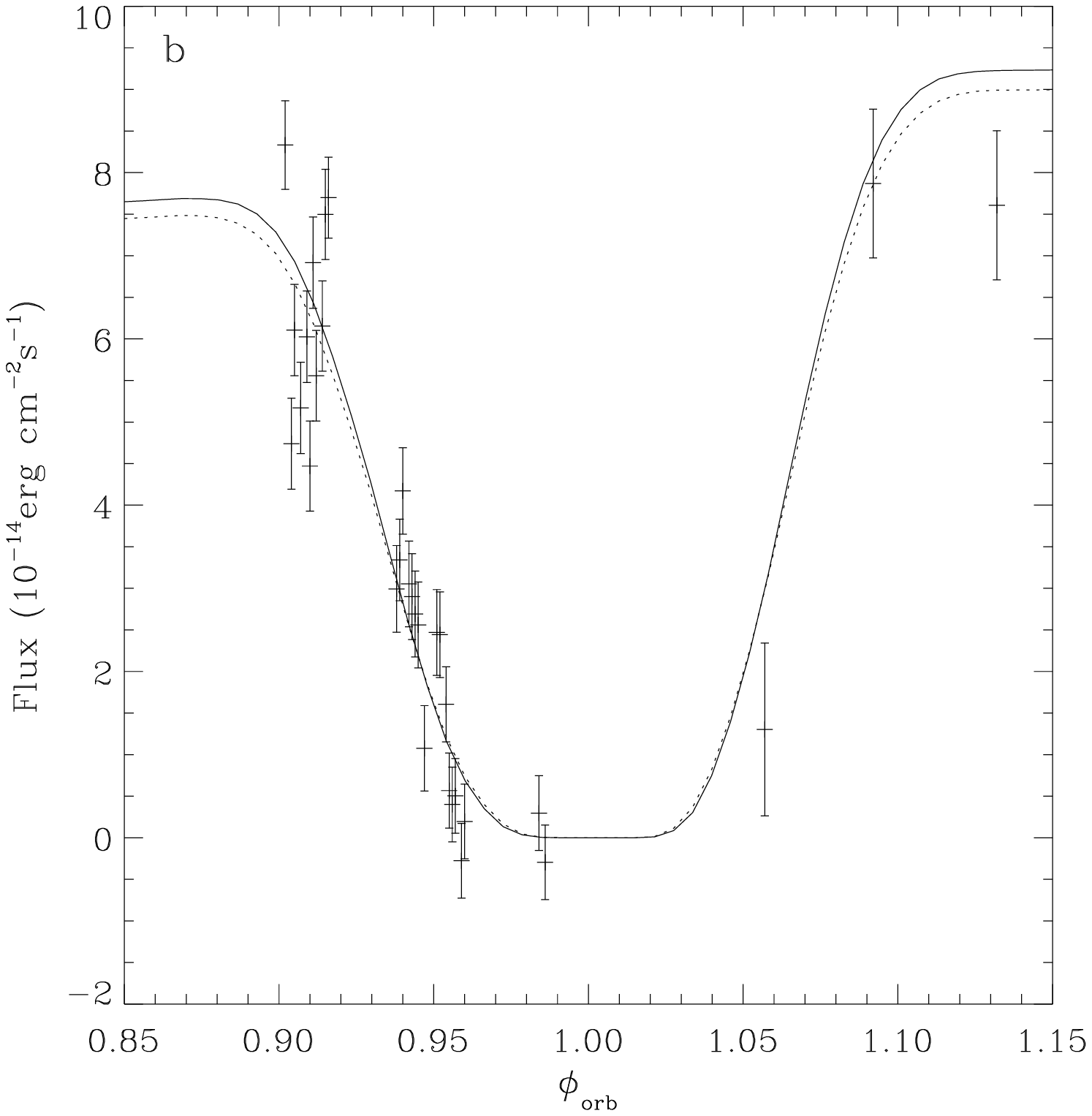}}
\caption{}
\end{figure}

\begin{figure}
\centerline{\epsfxsize=6in\epsfbox{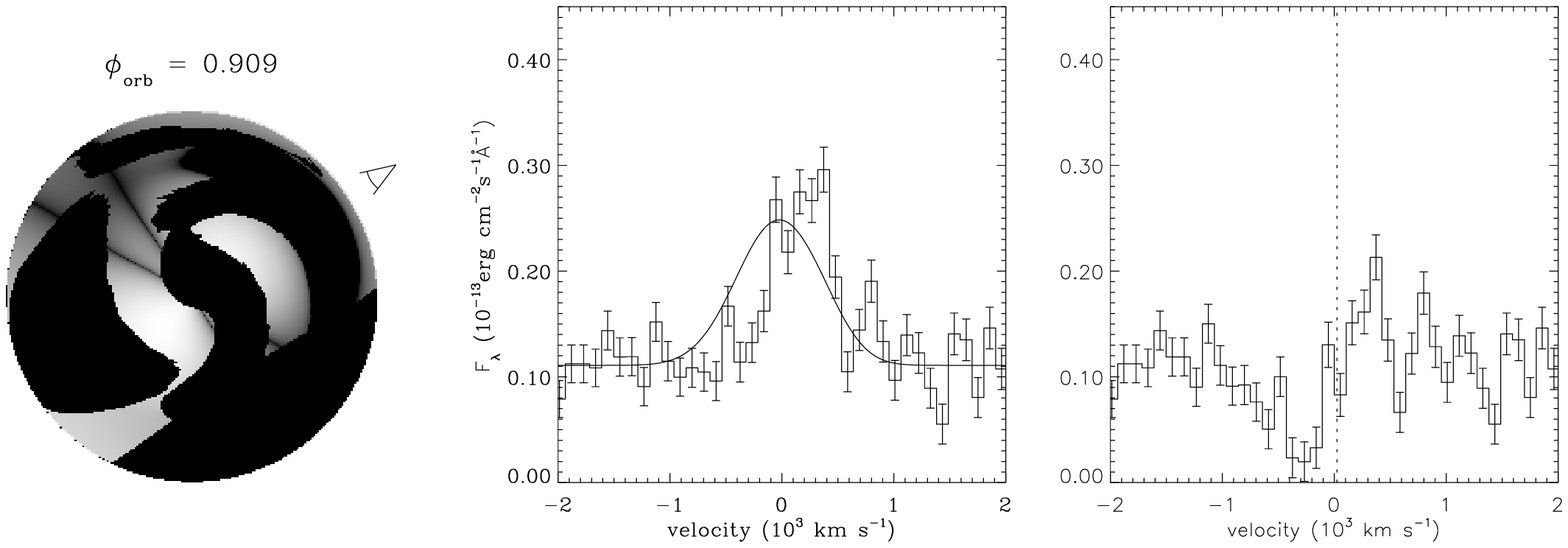}}
\centerline{\epsfxsize=6in\epsfbox{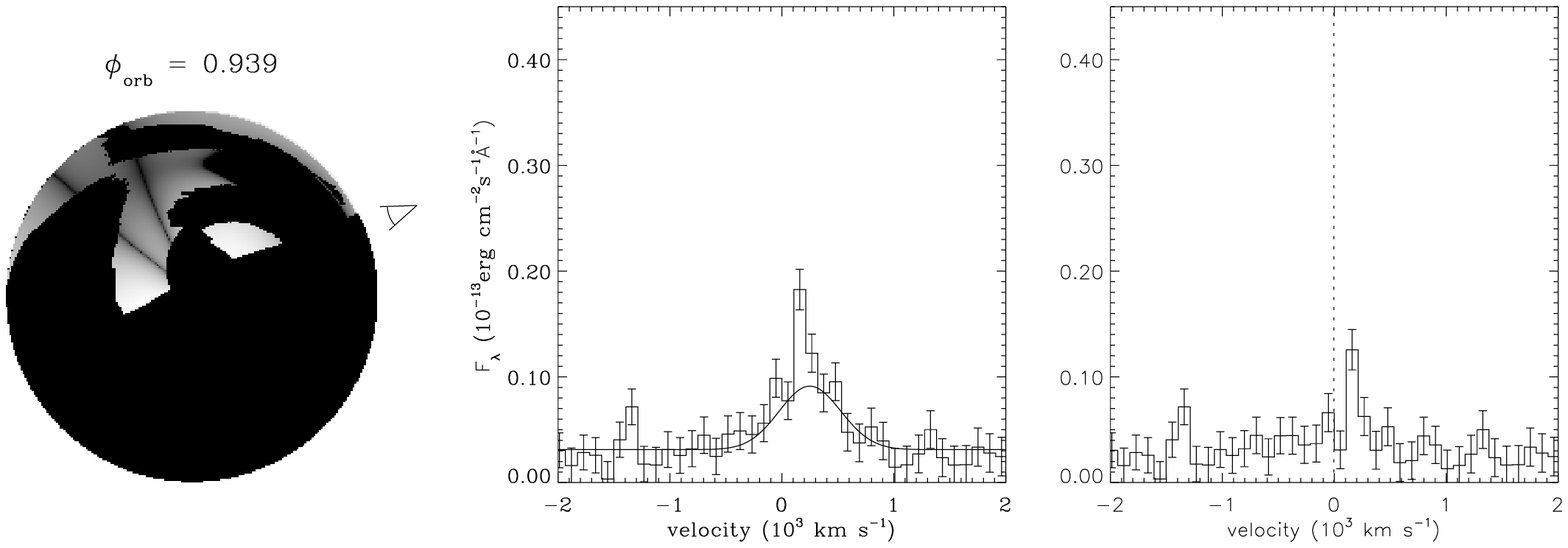}}
\centerline{\epsfxsize=6in\epsfbox{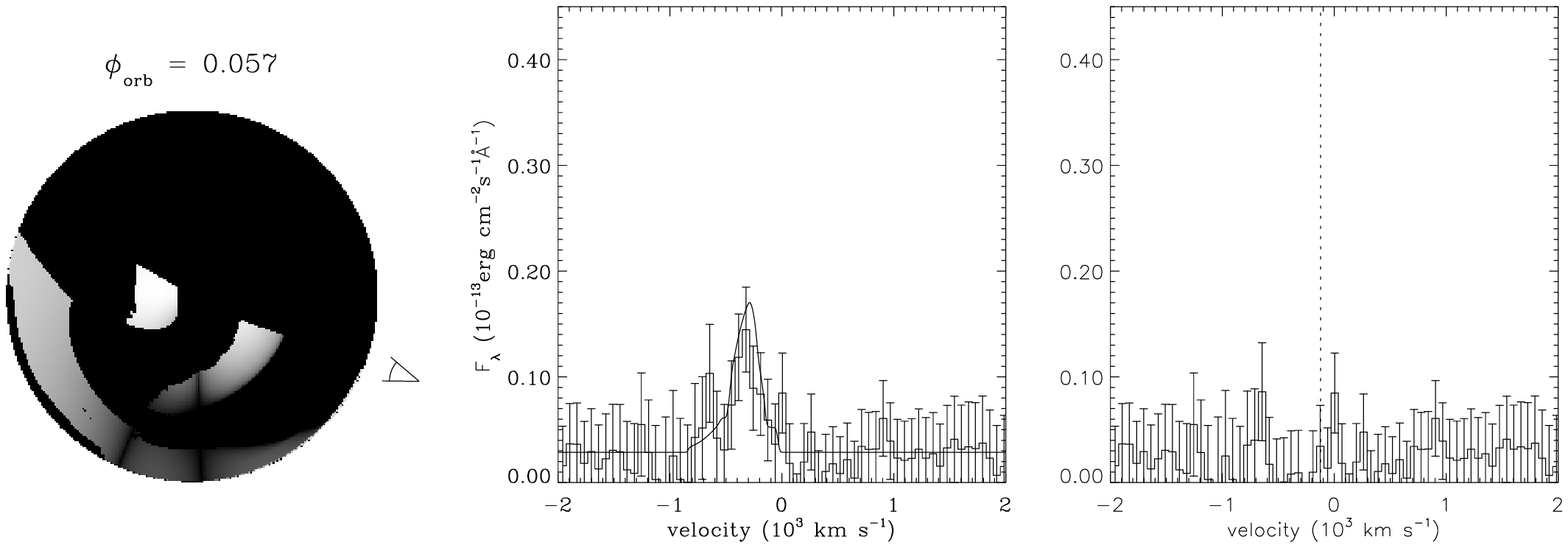}}
\centerline{\epsfxsize=6in\epsfbox{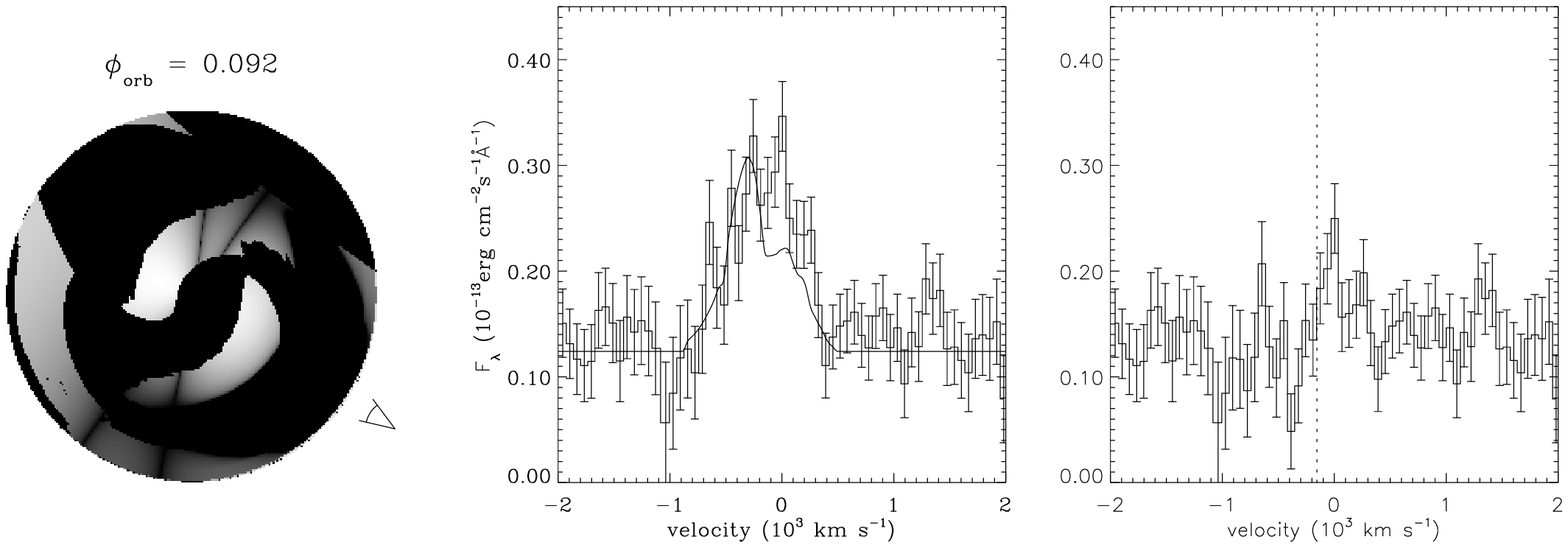}}
\caption{}
\end{figure}

\begin{figure}
\centerline{\epsfxsize=6in\epsfbox{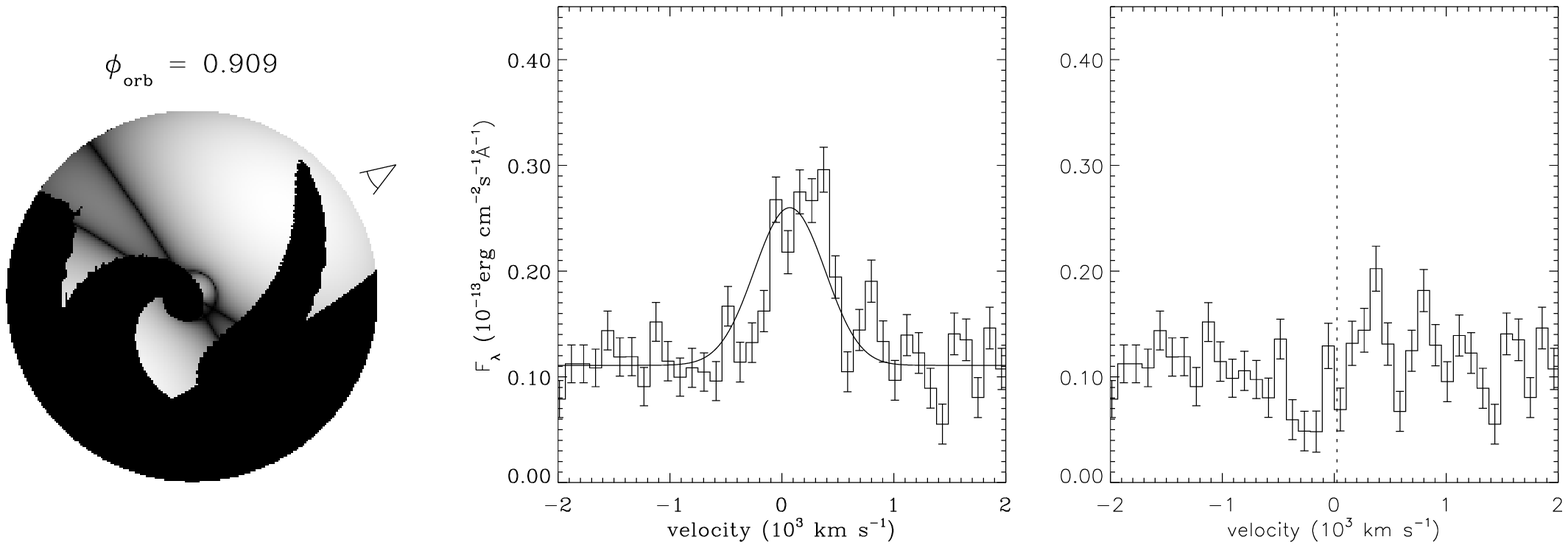}}
\centerline{\epsfxsize=6in\epsfbox{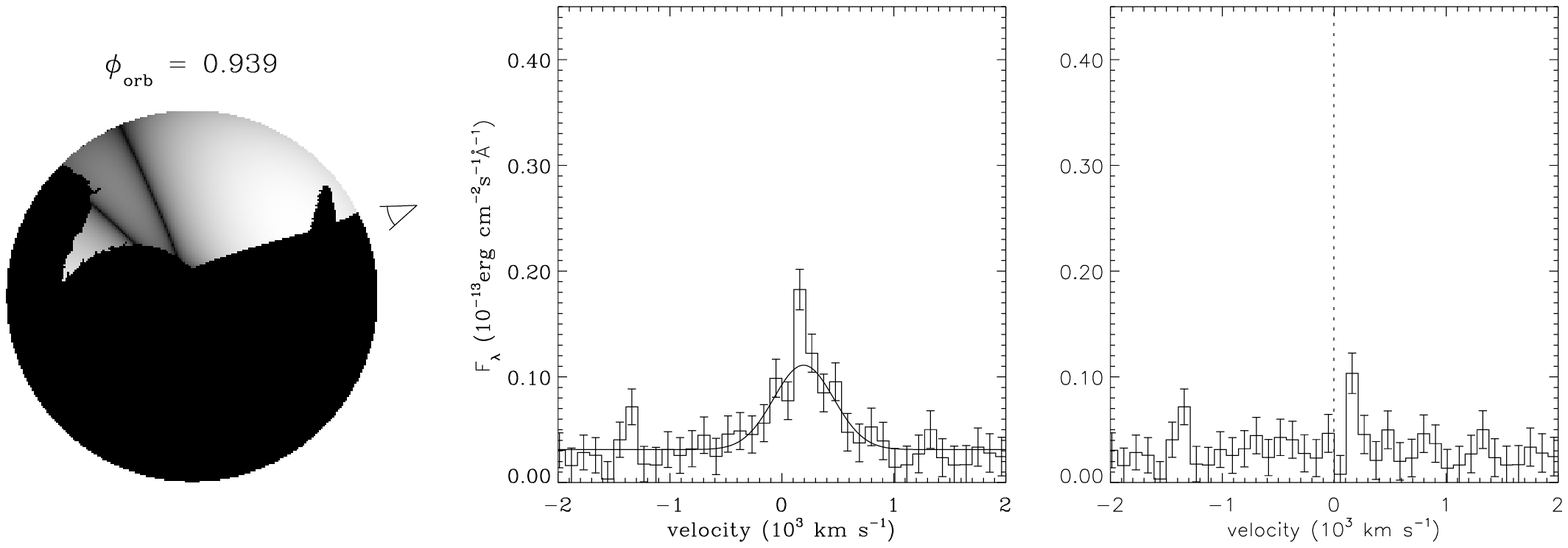}}
\centerline{\epsfxsize=6in\epsfbox{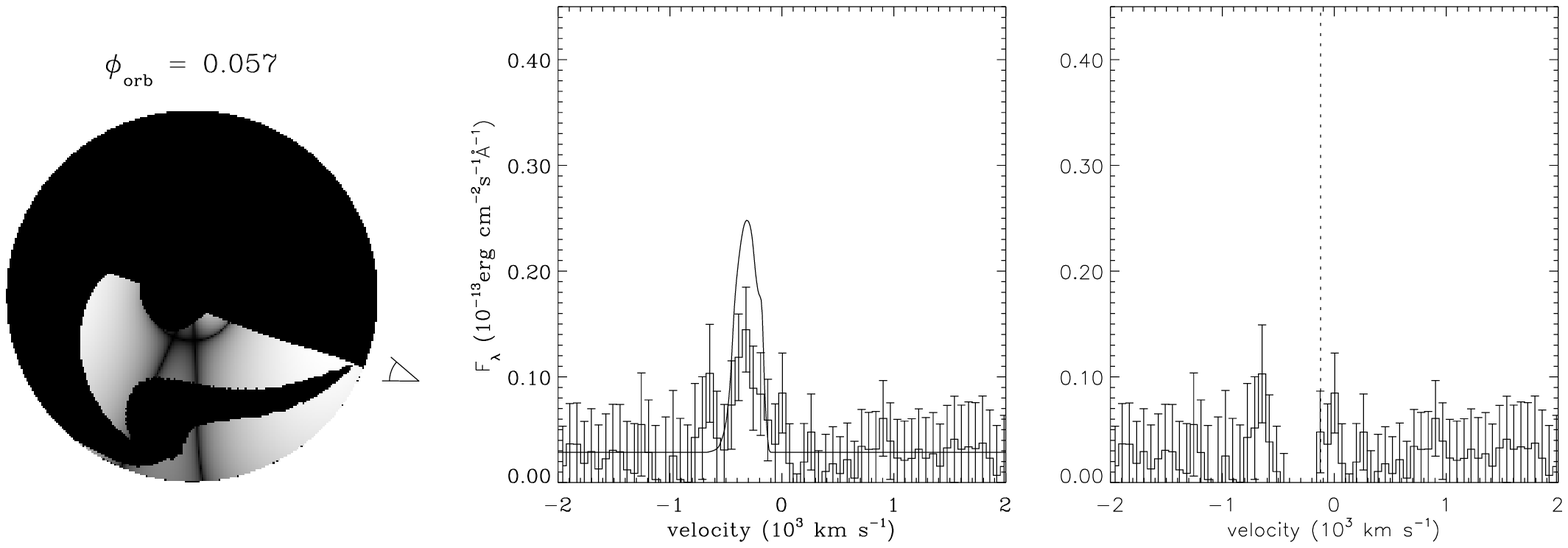}}
\centerline{\epsfxsize=6in\epsfbox{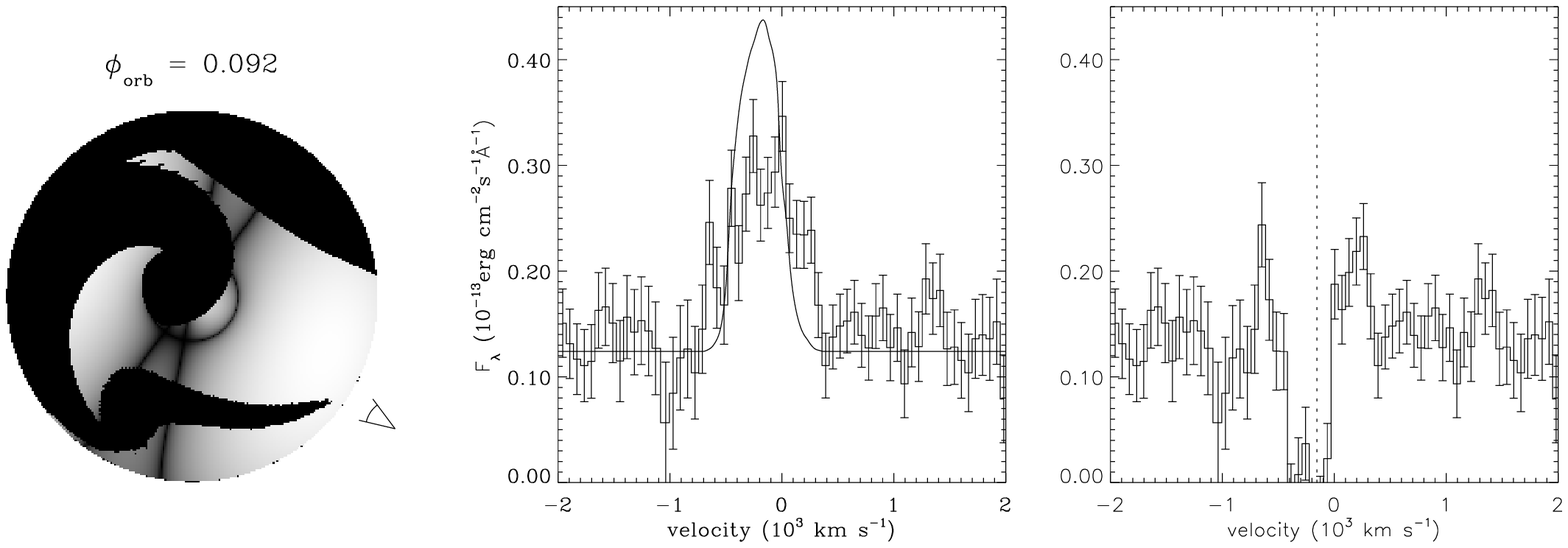}}
\caption{}
\end{figure}

\end{document}